\documentclass[%
 reprint,
 amsmath,amssymb,
 aps,
prb,
floatfix,
]{revtex4-2}

\usepackage{graphicx}
\usepackage{dcolumn}
\usepackage{bm}
\usepackage{svg}
\usepackage{subfigure}
\usepackage{tabularx}
\usepackage{multirow}
\usepackage{physics}
\usepackage{float}

\begin{document}

\preprint{APS/123-QED} 

\title{Modeling semiconductor spin qubits and their charge noise environment for quantum gate fidelity estimation }

\author{M.Mohamed El Kordy Shehata}
 \altaffiliation[Also at ]{KU Leuven Department of Physics and Astronomy.}
 \author{George Simion}%
  \email{george.simion@imec.be}
  \author{Ruoyu Li}
  \author{Fahd A. Mohiyaddin}
  \author{Danny Wan}
  \author{Massimo Mongillo}
  \author{Bogdan Govoreanu}
  \author{Iuliana Radu}
  \author{Kristiaan De Greve}
  	\altaffiliation[Also at]{KU Leuven Department of Electrical Engineering}
  \author{Pol Van Dorpe}
    \altaffiliation[Also at ]{KU Leuven Department of Physics and Astronomy.}
 \affiliation{
IMEC, Kapeldreef 75, 3001 Leuven, Belgium
 }
 
\date{\today}

\begin{abstract}
The spin of an electron confined in semiconductor quantum dots is currently a promising candidate for quantum bit (qubit) implementations. Taking advantage of the existing CMOS integration technologies, such devices can offer a platform for large scale quantum computation. However, a quantum mechanical framework bridging a device's physical design and operational parameters to the qubit’s energy space is lacking. Furthermore, the spin to charge coupling introduced by intrinsic or induced Spin-Orbit-Interaction (SOI) exposes the qubits to charge noise compromising their coherence properties and inducing quantum gate errors. We present here a co-modeling framework for double quantum dot (DQD) devices and their charge noise environments. We use a combination of an electrostatic potential solver, full configuration interaction quantum mechanical methods and two-level-fluctuator models to study the quantum gate performance in realistic device designs and operation conditions. We utilize the developed models together alongside the single electron solutions of the quantum dots to simulate one- and two- qubit gates in the presence of charge noise. We find an inverse correlation between quantum gate errors and quantum dot confinement frequencies. We calculate X-gate fidelities $> 97\%$ in the simulated Si-MOS devices at a typical TLF density of $10^{11}$ $\mathrm{cm^{-2}}$. We also find that exchange driven two-qubit SWAP gates show higher sensitivity to charge noise with fidelities down to $91 \%$ in the presence of the same density of TLFs. We further investigate the one- and two- qubit gate fidelities at different TLF densities. We find that given the small size of the quantum dots, sensitivity of a quantum gate to the distance between the noise sources and the quantum dot creates a strong variability in the quantum gate fidelities which can compromise the device yields in scaled qubit technologies.
\end{abstract}

\maketitle



\section{\label{intro} Introduction}

The intrinsic two-level nature of an electron's spin besides its relatively weak coupling to its environment promotes it as an attractive building block for quantum information. Gate defined quantum dots in silicon Metal-Oxide-Semiconductor (MOS) devices \cite{veldhorst_addressable_2014,veldhorst_two-qubit_2015,fogarty_integrated_2018,kim_low-disorder_2019,zwerver_qubits_2021} or in Si/SiGe \cite{kawakami_electrical_2014,yoneda_quantum-dot_2018} and GaAs/AlGaAs \cite{petta_coherent_2005,koppens_driven_2006,medford_self-consistent_2013} heterostructures offer a platform for the physical implementation of spin-based quantum bits (qubits). Benefiting from the well established CMOS fabrication techniques, spin qubits in silicon open the prospect for large scale quantum computers.  Furthermore, development of technologies producing isotopically purified Si-28 wafers \cite{ager_high-purity_2005}, in addition to silicon's relatively weak Spin-Orbit Interaction (SOI), allowed Si-based spin qubits to circumvent the main obstacles facing its GaAs/AlGaAs counterparts. 

As proposed by Loss and Divincenzo in \cite{loss_quantum_1998}, controlled exchange interaction can drive two-qubit gates in spin based implementations. In quantum dot implementations, the exchange interaction can be electrically controlled using gate voltages modulating the overlap between the two electron wave functions. While spin qubits in silicon have been experimentally demonstrated with one-qubit gate fidelities larger than 99.9\% \cite{yoneda_quantum-dot_2018,veldhorst_addressable_2014}, two qubit gates performance is still limited \cite{veldhorst_two-qubit_2015,huang_yang_FidleityBench_2019,watson_philips_Programmable,zajac_Resonantly_2018}. Despite the relatively large coherence times of spin qubits in silicon, the two-qubit gates' performance is limited by its operation speed, which depends on the value of exchange, in addition to sensitivity of exchange interaction to charge noise. Such limitations can be further pronounced in large scale implementations of spin qubits in silicon. Accordingly, a quantum mechanical modeling framework for coupled quantum dots can estimate the exchange interaction, its dependence of operation parameters such as voltage-controlled tunnel couplings and energy detuning and more importantly its sensitivity to noise that directly affects the two-qubit fidelities \cite{huang_spin_2018,nielsen_conguration_nodate,kim_coupled_2008,jiang_coulomb_2013}. 

In this work, we present a versatile full configuration interaction (FCI) quantum mechanical model that estimates exchange interaction based on electrostatic potentials simulated from realistic devices, in contrast to oversimplified parabolic potentials that were considered in previous works \cite{nielsen_conguration_nodate,kim_coupled_2008,jiang_coulomb_2013}. The model offers the possibility of studying the effect of different device parameters on exchange interaction. Furthermore, the model allows the inclusion of any external potentials to the system which opens the door to study the effect of microscopic noise sources, individually or collectively, on the qubits and quantum gate fidelities.
	
\begin{figure}[t]
    \centering
    \includegraphics{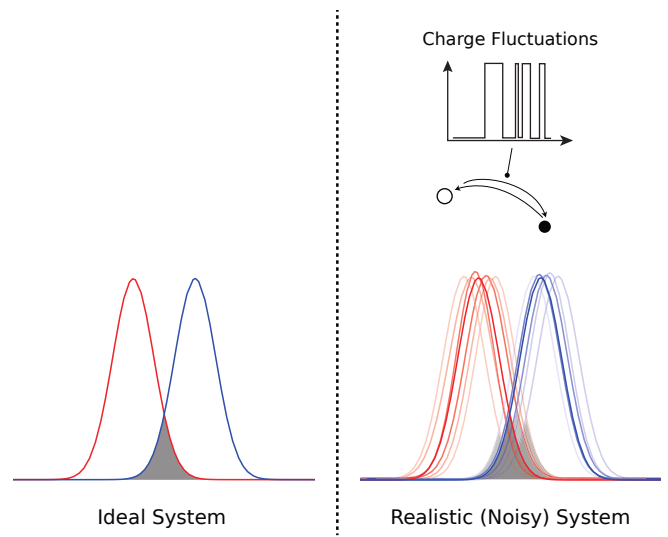}
    \makeatletter\long\def\@ifdim#1#2#3{#2}\makeatother
    \caption{\small An illustration of the wave functions of two electrons in two quantum dots and their overlap in an ideal system (left panel) and a realistic system with charge noise (right panel). }
    \label{fig:Illustration}
\end{figure}

Charge noise has been reportedly suggested to be the limiting factor of coherence behavior of spin qubit implementations in silicon such as Si-MOS devices and SiGe/Si/SiGe heterostructures. Such a hypothesis is motivated by the observed 1/f -like spectra extracted from spin qubits noise spectroscopy methods \cite{cywinski_2014,szankowski_2017,yoneda_quantum-dot_2018,chan_assessment_2018,connors_charge-noise_2022}. 1/f noise spectra are historically associated with charge noise in electronic devices and have been extensively studied \cite{machlup_noise_1954,dutta_low-frequency_1981,fleetwood_1f_2015}. While intrinsic SOI in silicon is relatively weak, the reported charge noise-limited decoherence behavior of spin qubits suggest some spin to charge coupling introduced in the system which may be a result of stray magnetic field gradients, spatial modulations of the g-factor or SOI at interfaces. Such mechanisms can cause dephasing in the qubit in the presence of charge noise. Furthermore, such coupling has become an essential tool for implementing fast electrically controlled one- and two-qubit gates \cite{pioro_2008,Brunner_2011,Giavaras_2019,Giavaras_2019_2}. As discussed earlier, two-qubit gates rely on controlled exchange interactions between two spins in two quantum dots. The electrostatically mediated exchange interactions render such systems prone to errors due to charge noise. In a simple illustration, Fig.\ref{fig:Illustration} describes the modulation of the electron wave function due to charge noise fluctuations. Such modulation can cause qubit frequency shift in the presence of intrinsic or artificial SOI, effectively causing dephasing. Furthermore it is also reflected in the overlap between two electrons' wave functions which in turn appears as fluctuations in the exchange interaction, inducing two qubit gate errors.. In this paper we develop a simple charge noise model based on the electrostatic fluctuations induced by its microscopic sources represented by Two-Level-Fluctuators (TLFs).

The paper is organized as follows. In Sec.\ref{app:DQD Model} a highly optimized full configuration interaction model is described and demonstrated on realistic device designs of laterally coupled double quantum dots (DQD). In Sec.\ref{sec:Theoretical Description} a theoretical description of 1/f noise is outlined in addition to a discussion on the sources of such noise and their electrostatic potentials. In Sec.\ref{sec:Simulations}, Monte-Carlo type simulations of charge noise spectra  for three types of TLFs are shown and compared to experimentally reported figures. In Sec.\ref{sec:Quantum Gate Fidelities}, both models are used in conjunction with each other alongside time-evolution simulations to study the effect of charge noise on quantum gate fidelities. The summary of our findings is reported in Sec.\ref{sec:Conclusions} of this paper.

\section{Double Quantum Dot Model}
\label{app:DQD Model}
\begin{figure}
    \centering
    \includegraphics{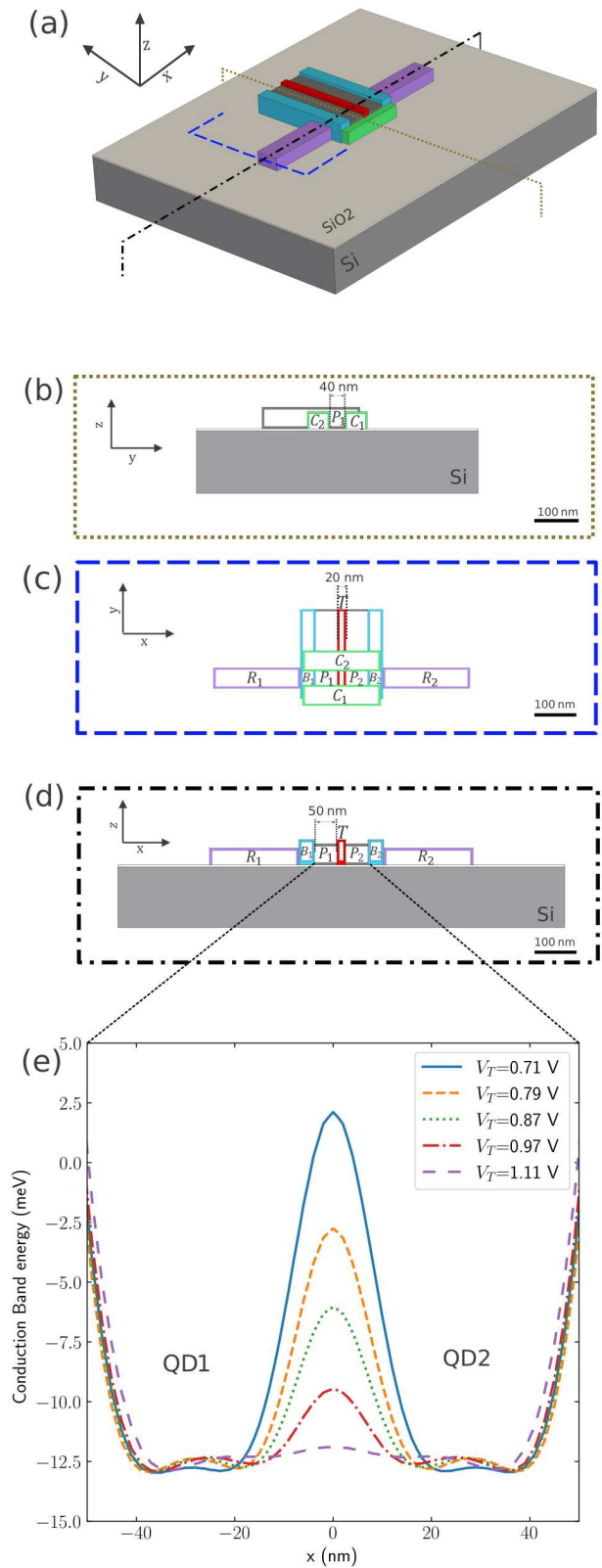}
    \caption{(a) 3D Geometry of the simulated DQD device with (b) y-z ,(c) x-y and (d) x-z sections.The quantum dots are defined under the $\mathrm{P_{1,2} }$ gates which are 50 nm x 40 nm and separated by a 20 nm wide tunnel (T) gate.(e) shows 1D slices of simulated electrostatic potentials of the device at different tunnel gate ($V_T)$ voltages.}
    \label{fig:3D}
\end{figure}

A most basic building block for spin-based quantum computing in quantum dots is a double quantum dot (DQD) system. In a low coupling regime, electrons in each of the dots can be addressed and controlled as single qubits. In a high coupling regime, the exchange interaction between the two electron spins drives two-qubit gates required for computations. Coupling between the dots is controlled either via tunnel coupling (t) or energy detuning ($\mathrm{\epsilon_d})$. Here we develop a quantum mechanical model to estimate the energy spectrum of two electrons in a DQD system and is demonstrated on the device shown in Fig.\ref{fig:3D} at different spacings between the quantum dots and different operation points. Here we assume a symmetric operation \cite{reed2016} of the DQD, where the exchange interaction is controlled by tuning the tunnel coupling. However, the modeling framework allows for tuning both tunnel coupling and energy detuning.

The device in Fig.\ref{fig:3D} is a Si-MOS DQD system with three metal (TiN) gate layers separated by 5 nm of $\mathrm{SiO_2}$.  The first gate layer includes the confinement gates,  $\mathrm{C_{1,2}}$, and the reservoir gates, $\mathrm{R_{1,2}}$, under which Ohmic regions are defined with phosphorus concentration of $\mathrm{10^{18} cm^{-2}}$. $\mathrm{C_{1,2}}$ act as screening gates and can provide extra confinement for the quantum dots along the y-direction while  $\mathrm{R_{1,2}}$ are used to form a 2D electron gases (2DEG) used as electron reservoirs. The second gate layer includes the plunger gates, $\mathrm{P_{1,2}}$, under which the quantum dots are defined. The third gate layer includes the tunnel, T, gate and the barrier, $\mathrm{B_{1,2}}$, gates which control the tunnel coupling between the two dots and between the dots and the reservoirs respectively. In this device geometry the quantum dot size is 50 nm x 40 nm. Note that the tunnel gate width in the device shown is 20 nm, however, later in this section calculations are done on the device with varying the tunnel gate width from 20 nm to 50 nm which corresponds to inter-dot distance varying from 70 nm to 100 nm. In this geometry we define the x-y plane as the in-plane surface and the z-axis as the out-of-plane direction. In this work, 3D electrostatic potentials, obtained from Technology Computer Aided Design (TCAD) simulations performed with \textit{Sentaurus} software on realistic double quantum dot devices, are used to describe the devices potential profile. 1D slices of the extracted TCAD electrostatic potentials at different tunnel gate voltages are shown in Fig.\ref{fig:3D}(e). It can be seen in the figure that there is a "bump" in center of the potential of each of the dots. We believe that those are numerical artifacts of the electrostatic simulator since it allows for fractional charge densities which are here formed in the middle of the dots, slightly raising the electrostatic potentials. Albeit, we believe such "bumps" have no major effect on the accuracy of the results presented here or the trends observed.

\begin{figure}[h]
\includegraphics{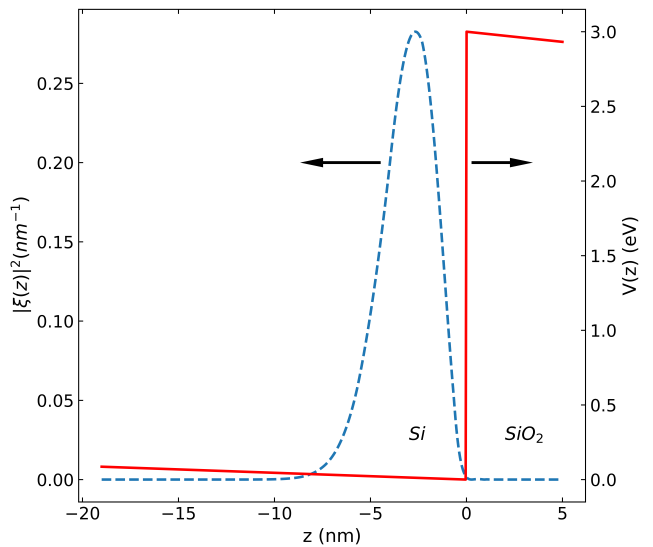}
\caption{The numerically calculated out-of-plane ground state probability distribution (blue dashed line) over a sketch of the out-of-plane potential (red solid line). }
 \label{fig:OutOfPlane} 
\end{figure}

In the simulations presented here, the P gates are set to 0.9765V, the R gates to 2.5V, the T gate is varied as shown in Fig.\ref{fig:3D}(e) and all other gates are set to 0V. At this set of voltages, the DQD electrons are accumulated approximately 3 nm below the $\mathrm{Si/SiO_2}$ interface (see Fig.\ref{fig:OutOfPlane}) with 2DEGs formed under the reservoir gates. The tunnel barriers between the quantum dots and the electron reservoirs (2DEGs) are approximately 100 meV high, hence, electrons are safely assumed to be confined within the quantum dots with no components inside the reservoir.

Estimating the two-qubit gate-driving exchange interaction requires solving the two-electron Hamiltonian (Eq.\ref{eq:twoelectronH}) which in turn governs the system's evolution in time as described by Schrodinger's equation (Eq.\ref{Schrodinger}).
\begin{equation}
\begin{split}
   \hat{H}=\sum_{\zeta}^{x,y,z}\sum_{i}^{1,2}&\frac{-\hbar^2}{2m_{\zeta}^*}\frac{\partial^2}{\partial\zeta_i^2}+V(\mathbf{r_1})+V(\mathbf{r_2})\\
    +&g\mu_b \mathbf{B.S} +\frac{e^2}{4\pi\epsilon\mathbf{|r_1-r_2|}},
    \end{split}
\label{eq:twoelectronH}
\end{equation}
\begin{equation}
 i\hbar\frac{\partial}{\partial t}  \ket{\mathbf{\Psi}}= \hat{H} \ket{\mathbf{\Psi}},
\label{Schrodinger}
\end{equation}
where $\hat{H}$ is the Hamiltonian describing the energy of two interacting conduction band electrons within an effective mass approximation. The first term is the kinetic energies of the two electrons denoted 1 and 2, with ($m_{x(y)}^* =0.19m_0$ in-plane  \& $m_z^*=0.98m_0$ out-of plane) \cite{sze_ng_1981} the effective mass of the conduction band electrons. $V(\mathbf{r_{1,2}})$ are functions describing the gate-defined potential landscape in which the two electrons reside with $\mathbf{r_{1,2}}$ being the spatial coordinates of the two electrons. $g\mu_b \mathbf{B.S}$ is the Zeeman term due to the external magnetic field ($\mathbf{B}$),  needed to create a two-level spin system, with $g$, the electron g-factor, $\mu_b$, the Bohr magneton, and $\mathbf{S}$, the spin operator. For estimating the exchange interaction, the magnetic field is assumed to be constant over the two dots and is only included to lift the degeneracy between the triplet states, magnetic field gradients are included later in the text in treatment of one- and two- qubit gates. Finally, the last term is the two-electron Coulomb interaction with $\epsilon$, the material permitivity. 

In the literature \cite{nielsen_conguration_nodate,kim_coupled_2008,jiang_coulomb_2013,White_2018}, simplified potentials are often used to describe $V(\mathbf{r})$. Typically a parabolic potential for a single quantum dot or two such potentials next to each other with a Gaussian barrier in between them for a double quantum dot system are considered. While parabolic potentials can accurately describe single quantum dots at low energies, they do not reflect the dependence of the electrostatic potential on the device design and applied voltages, especially when dealing with more than one dot such as in a double quantum dot (DQD) system. Furthermore, 2-dimensional treatment of such systems is usually adopted, reducing computational time and complexities at the expense of accuracy. 
As described above, the model presented here takes numerical 3D potentials from electrostatic simulations of realistic devices as an input and estimates the systems energy spectrum. This allows for its applicability to different quantum dot topologies, such as 2D arrays. Such potentials offer a more realistic view of the systems in study while offering insights on their sensitivities to actual physical parameters such as gate sizes, critical distances, oxide thickness and applied voltages. 

The two-level states encoding a qubit are usually found within a much larger energy spectrum of the hosting system. For instance, in addition to its spin, an electron in gate defined quantum dots essentially possesses an orbital degree of freedom defining its position in space alongside a valley degree of freedom in some materials such as silicon and germanium. The energies of the single electron part of the Hamiltonian $\hat{H}$ can be calculated in a straight forward approach using Schrodinger-Poisson solvers. On the other hand, estimating the exchange interaction requires calculating the two-electron energy spectrum of the full Hamiltonian $\hat{H}$ where it is defined as the energy splitting between the lowest un-polarized singlet ($S_0$) and triplet ($\mathrm{T_0}$) states.

As in any scenario involving multiple electrons, the Coulomb term in Eq.\ref{eq:twoelectronH} presents a computational challenge by introducing electron correlations in addition to its divergence at $r_1=r_2$. The full configuration interaction (FCI) method solves the Hamiltonian in Eq.\ref{eq:twoelectronH} by diagonalizing its matrix representation over an extended basis set. The exact solution is attained as the number of basis elements goes to infinity. Albeit, with an appropriate choice the basis set, acceptable convergence can be reached at a practical number of basis set elements.

A first step towards finding an appropriate basis set for solving the two-electron Hamiltonian, is finding the single electron solutions of the acquired 3D potentials. Here we assume the in-plane (x-y) and out-of-plane (z) potentials are independent leading to a separation of variables for the single electron wave function ($ \ket{\mathbf{\Psi_s}}$) such that 
\begin{equation}
 \ket{\mathbf{\Psi_s}}=\Psi_s(x,y,z)= \psi_{n^x}(x)\psi_{n^y}(y)\xi_{n^{z}}(z),
\label{eq_singlewavefunction}
\end{equation}
with $\mathrm{n^{i}}$ is the nth eigen-function of the potential in the i-direction where $\mathrm{i \in {x,y,z}}$. Solving the out-of-plane (z), single-electron part of the Hamiltonian, $\hat{H}$ in Eq.\ref{eq:twoelectronH} numerically, we find the ground state energy $\approx$ 20 meV. The z-direction ground state squared wave function is shown in Fig.\ref{fig:OutOfPlane}, showing a strong confinement in that direction with the quantum dots formed approximately 3 nm below from the $\mathrm{Si/SiO_2}$ interface. Such a wave function closely resembles an Airy function, which is not surprising given the expected triangular potential profile in the z-direction as sketched in Fig.\ref{fig:OutOfPlane}. 

\begin{figure}[h]
\includegraphics{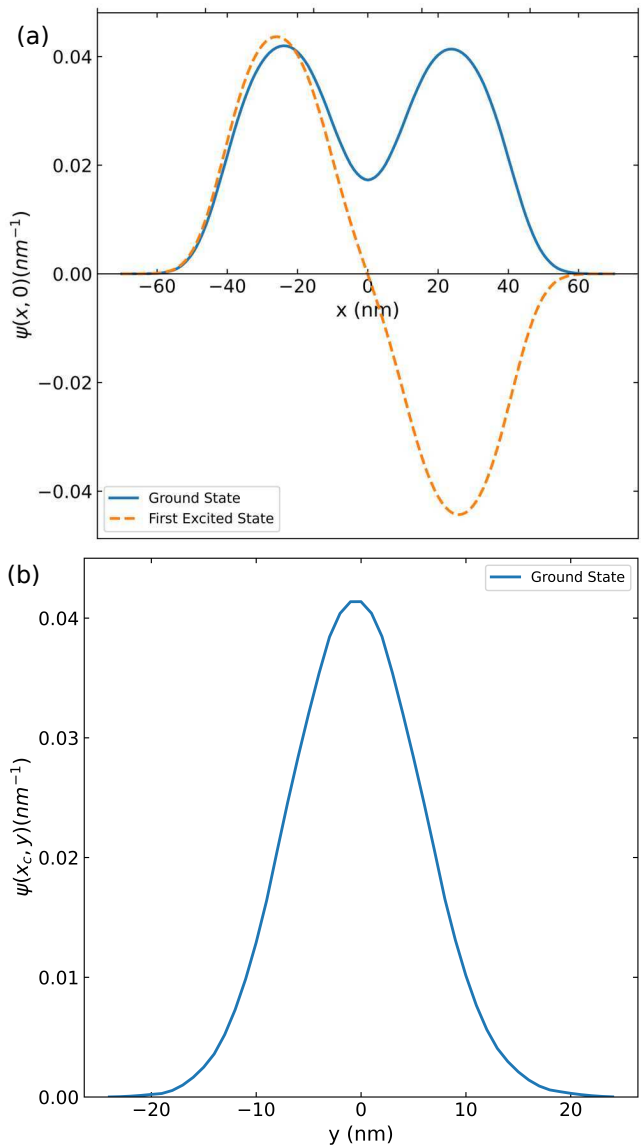}
\caption{The numerically calculated in-plane wave functions of the ground and first excited states. (a) shows the wave functions in the x-direction at y =0 . Both states have the same wave function in the y-direction shown in (b) as the ground state in the y-direction with x set to $\mathrm{x_c\approx \pm 25 nm}$, the center of any of the two dots }
 \label{fig:InPlane} 
\end{figure}

Solving the in-plane (x-y), single-electron part of the Hamiltonian, $\hat{H}$ in Eq.\ref{eq:twoelectronH} numerically, we find the first two eigen-functions shown in Fig.\ref{fig:InPlane}. In the y-direction the solution perfectly fits to the harmonic oscillator ground state centered at the origin. On the other hand, in the x-direction the solutions resemble symmetric and anti-symmetric combinations of harmonic oscillator ground states centered at each of the quantum dots. The single-electron solutions presented here are found at tunnel gate voltage ($\mathrm{V_T}$) of 0.97 volts which corresponds to a tunnel barrier height of approximately 3 meV and tunnel coupling of 33.92 GHz.

\begin{figure}[hb]
\includegraphics{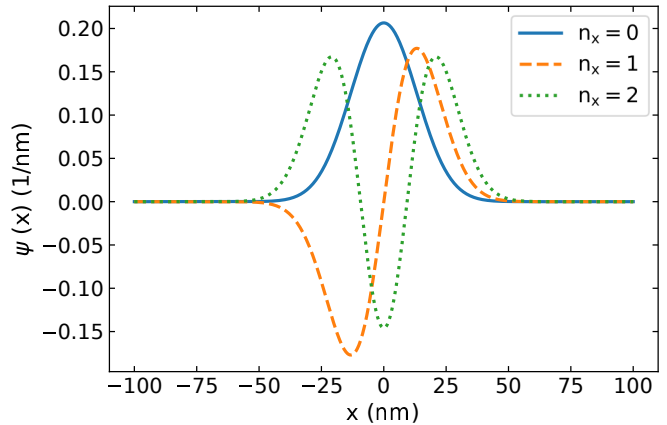}
\caption{1D plots of the first three wave functions in the central basis set}
 \label{fig:CentralBasis} 
\end{figure}

Accordingly, a natural choice of basis for solving the two-electron wave function would be the triangular potential eigen-function in the z-direction, harmonic oscillator eigen-functions, centered at the origin, for the y-direction and finally harmonic oscillator eigen-functions centered at each of the two dots for the x-direction. However, such a choice of basis comes with the need for orthogonalization and the need to recalculate all one- and two-electron integrals with new device geometries and/or operational conditions. Instead, we chose a "central basis set" where harmonic orbitals are chosen to be centered in between the two quantum dots (x=y=0) (Fig.\ref{fig:CentralBasis}) and are described by 

\begin{equation}
\begin{split}
\psi_{n^x}(x)\psi_{n^y}(y) &=\frac{1}{\mathcal{N}} e^{\frac{-x^2}{2\lambda_x^2}} e^{\frac{-y^2}{2\lambda_y^2}} H_{n^{x}}(x/\lambda_x) H_{n^y}(y/\lambda_y)\\
&=\psi_{n}(x,y),
\end{split}
\label{eq:Hermite}    
\end{equation}
where $\mathrm{n={n^x+n^y}}$, $H_n$ is the nth Hermite polynomial, $\lambda_{x,y}$ are the radii of the wave function in the x and y direction and $\mathcal{N}$ is the normalization factor.
The use of a such "central basis set" rather than harmonic orbitals centered on the left and right dots reduces the complexity of the calculations by eliminating the need for orthogonalization. More importantly, such basis allow for a one time calculation of the Coulombic and kinetic integrals, drastically reducing the calculation time. Finally, the "central basis set" is independent of the underlying device potential,  providing versatility to extend to larger quantum dot systems.  The radii of the central harmonic orbitals ($\lambda_x , \lambda_y$) are estimated using the numerical single electron solutions of the in-plane potentials (Fig.\ref{fig:InPlane}) and are then used the scale the pre-calculated kinetic and Coulombic integrals.

To confirm the validity of the use of such basis, the single electron part of the Hamiltonian $\hat{H}$ was implemented as a matrix in the basis in Eq.\ref{eq_singlewavefunction}. Given the large energy spacing in the z-direction which is found to be larger than 15 meV, we chose to use a basis set of one ($\mathrm{n_z=0}$) element in this direction, keeping the numerical solution calculated earlier. We diagonalize the matrix with $\mathrm{n_x}$ up to 50 and $\mathrm{n_y}$ up to 4. We find the tunnel coupling, represented in the energy splitting between the first two eigen-functions, to match the values calculated from the TCAD simulations and the numerical calculations that followed. Hence, confirming the validity of the central basis set. Tunnel couplings calculated with only 25 basis set elements in the x-direction and only 1 in the y-direction were found to be within less than 1\% the ones calculated with more basis set elements (see App.\ref{app:MoreBasis}). This can be explained by the stronger confinement in the y-direction in addition to the need for a larger number of zero-centered basis set elements to represent wave functions at each of the dots. 
Nonetheless, the modeling framework presented here allows for the use of an arbitrary number of basis set elements in any of the three dimensions. Generally, quantum dots can demonstrate more symmetric confinement potentials especially in large scale implementations such as 2D array. In such case, the model presented here can easily accommodate such devices by changing the number of basis set elements accordingly. Some of the exchange calculations presented later in this section are repeated with more basis set elements in the z- and y- directions yielding very close results. Hence, we choose to work with a minimal basis set to optimize computational time.

Before proceeding with solving the full two electron Hamiltonian, it is important to note that the potential profile in Fig.\ref{fig:3D}(d) represents the silicon conduction band edge in the DQD area with the Fermi level as a reference at 0 eV. Therefore, all wave functions shown in Fig.\ref{fig:OutOfPlane},\ref{fig:InPlane} and described in Eq.\ref{eq_singlewavefunction},\ref{eq:Hermite} are envelope wave functions modulating the Bloch wave functions that reflect the underlying crystal potential. Silicon has 6 degenerate conduction band minima  (valleys) at $k_{\pm x,y,z}$.  Given the elliptical nature of such valleys, the effective masses are anisotropic, hence, degeneracy is lifted due to the strong confinement in the z-direction leaving only the two $k_{\pm z}$ valleys as the lowest valley states \cite{ando_fowler_stern_1982,culcer_2010}. $|k_{\pm z}|$ is approximately $0.85 \frac{2\pi}{a_0}$, with $a_0$ the lattice constant of Si \cite{SaraivaValley,saraiva_2009}. Using the same method proposed in Ref.\cite{SaraivaValley}, we calculate the valley coupling between the $k_{+z}$ and $k_{-z}$ valleys due to the $\mathrm{Si/SiO_2}$ interface potential to be approximately 65 $\mathrm{\mu eV}$ and a valley splitting of 130 $\mathrm{\mu eV}$. Therefore, with no degeneracy at the conduction band minima at the quantum dots, interference effects between the different valleys are suppressed for laterally (x-y) coupled QDs. Given that the in-plane potentials (Fig.\ref{fig:3D}(d)) and the eigen-functions used for the basis set are smooth over the scale of $a_0$, further interaction between valleys is negligible and the effective mass approximation is valid for silicon quantum dots \cite{culcer_2010,hada_eto_2005,Koiller_2001}. We further demonstrate the validity of the effective mass approximation in the calculations presented in App.\ref{app:EMA}.With that, we proceed to use the envelope wave functions described earlier as basis for finding a linear combination of harmonic orbitals (LCHO) \cite{puerto_gimenez_linear_2007} as solutions of the one- and two-electron wave functions in the DQD system. It is important to note here that despite the lifting of the valley degeneracy, interference of the two z-valleys can still be present for vertically coupled quantum dots and hence the full electron wave function that reflects the crystal symmetry would be required (see App.\ref{app:EMA}). Furthermore, as shown in Ref.\cite{tariq_2022}, any spatial modulations of the valley spectrum over the DQD structures can be detrimental to the exchange interaction. To elaborate, if the two dots do not have the same valley splitting and/or valley phase, the envelope wave functions would not suffice to describe the electrons' interaction. Therefore, in this work we assume that the valley structure is identical in the two dots. In App.\ref{app:Valley_Exchange_Steps} a first-order approximation of the effect of atomic steps at the $\mathrm{Si/SiO_2}$ interface on the valley-structure of the two dots and hence, the exchange is presented.

The utilization of a central basis set centered at the origin alongside the use of Hermite polynomials generating function, the Coulomb integrals of the form 
\begin{equation}
    \int dr_1dr_2\,\psi_{n_1}^{*}(r_1)\psi_{m_1}^*(r_2)\frac{e^2}{4\pi\epsilon|r_1-r_2|}\psi_{n_2}(r_1)\psi_{m_2}(r_2),
\label{eq:Coul}
\end{equation}
can be transformed into 
\begin{equation}
    \sum_{n}^{n1+n2}C_n\sum_{m}^{m1+m2}C_m\sum_{l,p}^{n+m}D_{nmlp} \int dr\, \psi_{p}(r) \frac{e^2}{4\pi\epsilon|r|},
\label{eq:TarnsCoul}
\end{equation}
where $r$ is the relative coordinate $r_1-r_2$ and the integration over the center of mass coordinate $R=\frac{r_1+r_2}{2}$ is performed analytically once and its result is absorbed in the summation coefficient $D_{nmlp}$. The coefficients of the summation ($C's \& D's$) are computed analytically once for a wave function of radius $\lambda=1$ nm and can be scaled to wave functions with other $\lambda$s with a simple multiplication and with no need for further calculations. Such a transformation reduces the number of unique Coulombic integrals from $N^4$ to $4N$ where $N$ is the number of central basis set elements used in the FCI calculation. A more detailed description of the transformation above can be found in App.\ref{app:Coeff}.

The work-flow of the model described above can be summarized in a few steps. First, electrostatic simulations are performed on the device structure to estimate the 3D potential. Second, in-plane ($x,y$) and out-of-plane ($z$) potentials are separated. Third, the Schrodinger equation is solved to find the single electron solutions out of which a numerical solution for the out-of-plane potential is acquired and the in-plane solutions are fit to Harmonic oscillator eigen-functions to estimate $\lambda_x$ and $\lambda_y$. Fourth, the single electron matrix from the first five terms of Eq.\ref{eq:twoelectronH} is computed in the central basis set (Eq.\ref{eq:Hermite}). Fifth, the two electron Coulomb matrix is computed using Eq.\ref{eq:Coul} \& Eq.\ref{eq:TarnsCoul} by scaling the coefficients of expansion using $\lambda_x$ and $\lambda_y$. Finally, the full matrix is diagonalized to estimate the exchange interaction. The exchange interaction is defined as the energy separation between the zero spin singlet ($\ket{S_0}$) and triplet ($\ket{T_0}$) states. Here $\ket{T_0}$ can be written as $\ket{(1,1)T_0}$ where 
$(N_L, N_R)$ denotes the number of electrons in the left and right dots respectively. On the other hand $\ket{S_0} \approx \ket{(1,1)S_0} + \theta \ket{(0,2)S_0}+\theta\ket{(2,0)S_0}$ where $\theta$ is the admixture factor with the double occupancy states due to the finite tunnel coupling \cite{huang_spin_2018}.

\begin{figure}[h]
\includegraphics{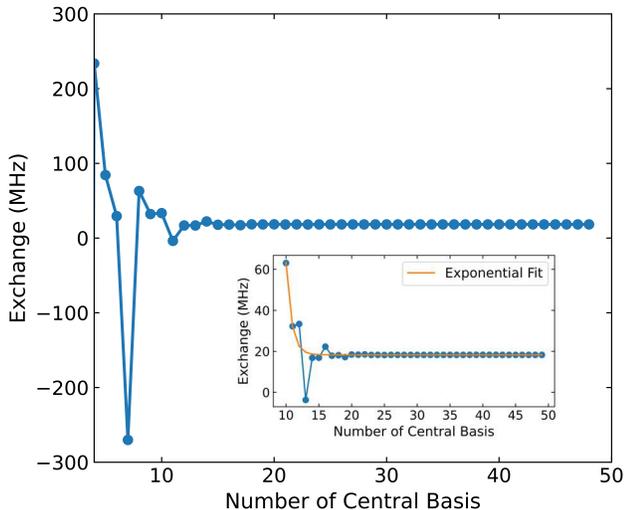}
\caption{Convergence of the exchange interaction by increasing the number of elements in the basis set. The inset shows an exponential fit to the convergence starting $\mathrm{n_x=10}$.
 \label{fig:Convergence} }
\end{figure}

Fig.\ref{fig:Convergence} shows the exchange interaction calculated using the full configuration interaction model at different number of basis set elements in the x-direction, with clear exponential convergence beyond $n_x=25$. In this FCI calculation performed here we assume the system is in its lowest valley state given the reported valley splitting values that are higher than the calculated exchange interactions in addition to the weak inter-valley Coulombic coupling \cite{SaraivaValley,Yang-orbitalandvalley,jiang_coulomb_2013,culcer_2010,tariq_2022}. However, in realistic devices, the dots' potential is randomized due interface roughness and steps. Such randomization is reflected in valley orbit coupling and hence the valley splitting which can differ between the dots allowing for inter-valley tunnelling \cite{culcer_2010,culcer2010}.  Furthermore, spin-valley mixing due to dipole interactions between the valley states have been observed and explained in Ref.\cite{yang2013}. The model presented here can easily adopt such perturbations by expanding the basis set to include the valley degree of freedom and the aforementioned perturbations can be introduced in the Hamiltonian matrix in a phenomenological approach.

\begin{figure}[h!]
\includegraphics{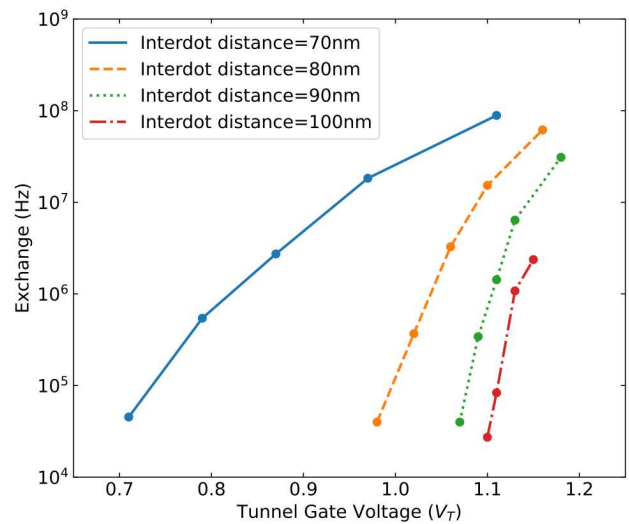}
\caption{Exchange interaction vs tunnel gate voltage calculated for DQD devices with different inter-dot distances. The plotted values are calculated using the full configuration interaction model. \label{fig:Exchanges} }
\end{figure}

Fig.\ref{fig:Exchanges} shows the exchange interaction calculated using the presented model at some tunnel gate voltages and for DQD devices with different inter-dot separations. The calculated exchange values fall in the MHz regime in agreement with experimentally reported values \cite{veldhorst_two-qubit_2015,zajac_Resonantly_2018,watson_philips_Programmable,AdamR2022,noiri2022fast,noiri2022shuttling}. Devices with the smallest inter-dot distances have shown tunability of the exchange interaction over a larger voltage range. Furthermore, it can be observed that at higher voltages the exchange interaction does not show an exponential dependence on the tunnel gate voltage, reflecting a similar non-exponential behavior of the tunnel coupling. Instead, signs of saturation appear indicating the hybridization of the two dots into a single larger dot, where the tunnel coupling approaches the orbital energy splitting of the single larger dot. Such a phenomena limits the extend to which the exchange interaction can be tuned. Such a transition naturally occurs at lower tunnel couplings for devices with larger inter-dot distance given that the orbital spacing of the hybridized single dot is smaller for bigger devices due to the weaker confinement. The exchange values shown in Fig.\ref{fig:Convergence} and Fig.\ref{fig:Exchanges} are calculated with a basis set of one element in both y and z directions for computational speed. However, the full 3D nature of the presented model allows for arbitrary number of basis set elements in any of the three dimensions as demonstrated in App.\ref{app:MoreBasis} for  a DQD device with 20 nm tunnel gate width and a tunnel gate voltage of 0.97V. The results shown in App.\ref{app:MoreBasis} show that the calculated exchange values presented here only show a slight change when more basis set elements are used in y- and z- directions. 

The quantum mechanical model described here offers a versatile framework for energy spectrum and exchange energy estimation for realistic devices using numerical 3D electrostatic potentials allowing for expansion to different quantum dot topologies such as 2D arrays. It also offers the flexibility to include any external potentials to the system. For instance, studying the effect of charge noise on the DQD system is achieved by including the electrostatic potentials from noise sources in the model. Modeling of charge noise from individual electrostatic fluctuations is developed and described in the next section.

\section{Noise Theoretical Description}
\label{sec:Theoretical Description}

In the context of semiconductor spin qubits, charge noise is essentially the electrostatic fluctuations in a quantum dot's environment. Such fluctuations modulate the quantum dot's confinement potential and hence the electron wave functions. In the presence of any spin to charge coupling mechanism, charge noise causes decoherence in the spin qubit and induces quantum gate errors. 
To represent charge noise, we adopt a general distributed trapping model \cite{dutta_low-frequency_1981} to describe the 1/f spectra. This approach describes charge noise as the collective effect of many individual two-level fluctuations caused by the hopping of charges around a device. The sources of such fluctuations are referred to as Two-Level Fluctuators (TLFs). We present a statistical description of the charge noise in Sec.\ref{subsec:TLF} and \ref{subsecRTSto1f}. We then examine the microscopic description of the electrostatics of the noise sources in Sec.\ref{subsecElectro}

\subsection{Two-Level Fluctuators (TLFs)}
\label{subsec:TLF}
\begin{figure}[b]
\includegraphics{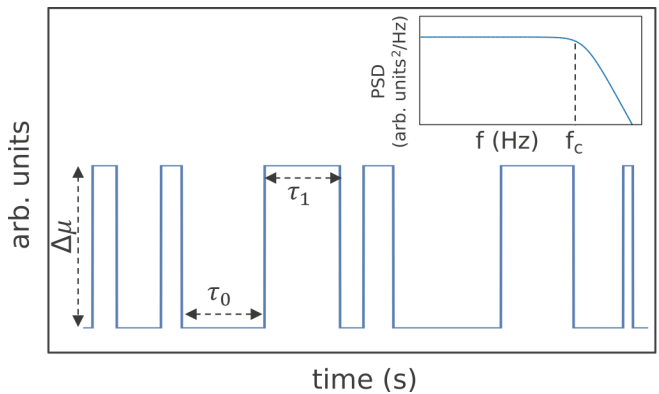}
\caption{\label{fig:OneRTN} A single Random Telegraph signal and its power spectrum (inset).}
\end{figure}

The signature effect of a TLF is observed as a bi-stable signal known as a Random Telegraph Signal (RTS) shown in Fig.\ref{fig:OneRTN}. The number of switching events in a RTS is described by a Poisson process and it can be characterized by three main parameters: the amplitude of the signal($\Delta \mu$), the average time spent in the high level ($\tau_1$) and in the low level ($\tau_0$). The power spectral density of a RTS is a Lorentzian as shown in the inset of Fig.\ref{fig:OneRTN} and is mathematically described by the formula \cite{machlup_noise_1954}

\begin{equation}
    PSD_{RTS}(f) = \frac{2 \Delta \mu^2}{(\tau_0 +\tau_1)((\frac{1}{\tau_0}+\frac{1}{\tau_1})^2 + (2\pi f)^2)}.
\label{eq_RTS_AS}
\end{equation}
As shown in Fig.\ref{fig:OneRTN}, the Lorentzian has a corner frequency $f_c =\frac{1}{\tau_0}+\frac{1}{\tau_1}$ . For simplicity, in this work symmetric TLFs will be assumed ($ \tau_0 = \tau_1 = \tau$ ) and the spectral density can be simplified to
\begin{equation}
    PSD_{RTS}(f) = \frac{ \Delta \mu^2 \tau}{4 + (2\pi f)^2 \tau^2},
\label{eq_RTS_S}
\end{equation}
where the times $\tau_0$ and $\tau_1$ are statistical parameters dependent on the kinetics of the underlying TLF while the amplitude ($\Delta \mu$) depends on the quantity under observation. For instance, decoherence and errors in quantum dot-based spin qubits are caused by fluctuations in the electrostatic potential whose amplitude, which essentially represents the coupling strength between the TLF and the qubit, has a $1/r$ dependence where $r$ is the distance between the fluctuator and the quantum dot. The switching time constants ($\tau_0$ \& $\tau_1$) have an exponential dependence on the tunnelling distance of the charge involved in the fluctuation. Under the assumption that the TLFs are not exchanging charges with the qubit device, the switching times are intrinsic to the fluctuators’ interactions with their environment and each other rather than with the quantum dot in study. 

\subsection{From RTS to 1/f Noise}
\label{subsecRTSto1f}
\begin{figure}
\includegraphics{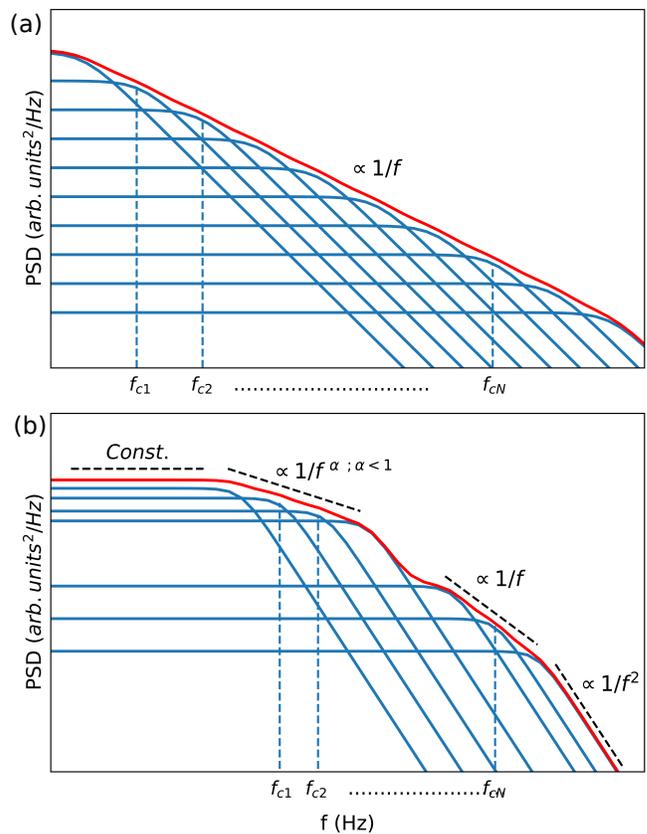}
\caption{\label{fig:RTNto1f} The superposition (Red Line) of multiple TLFs' random telegraph signals (Blue Lines) with (a) log-uniform distributed switching times giving a 1/f spectrum and (b) non-uniform distributions which can give $1/f^\alpha$ spectra with $\alpha<1$ for TLFs highly concentrated in a spatial and frequency region and $\alpha =2$ which is the Lorentzian tail of a single RTS spectrum. The blue lines are plotted using Eq.\ref{eq_RTS_S} and the red lines are their summation.}
\end{figure}

Under the assumption of zero cross-correlation between the individual switching events of different TLFs, the total power spectral density of a collection of fluctuators can be described as the sum of all individual RTS spectra in Eq.\ref{eq_RTS_S}. For a set of TLFs with the same amplitude and a log-uniform distribution of $\tau$ the well known 1/f noise spectrum can be observed as shown in Fig.\ref{fig:RTNto1f}(a). A uniform distribution of tunnelling distances in conjunction with the exponential dependence of tunnelling times on them can satisfy the later condition. However, as mentioned earlier, the TLF-qubit coupling strength is $1/r$ dependent, hence the distribution of the TLFs both in time and space contribute to the total noise sensed at the dot. Given the relatively small sizes of quantum dots, with radii typically in the tens of nanometers range, one would expect that the amplitude of the electrostatic fluctuations sensed by the quantum dot would be distributed over a wide range that would not satisfy an "all equal amplitude" approximation. Hence, the qubit-sensed noise would be dominated by the TLFs in close vicinity of the dot. In this context, 1/f noise is often described by $ A/f^\alpha$ where $\alpha$ usually varies from less than 1 up to 2 as shown in Fig.\ref{fig:RTNto1f}(b). Deviations from $A/f$ ($\alpha = 1$) in  nano-scale devices can be described by the small number of close TLFs and their nonuniform spatial and energetic distribution which can even sometimes lead to the observation of individual RTS Lorentzian spectra in noise measurements \cite{kafanov_charge_2008,connors_low-frequency_2019,struck_low-frequency_2020}. 

\subsection{Electrostatic Potentials}
\label{subsecElectro}
\begin{figure}
\includegraphics{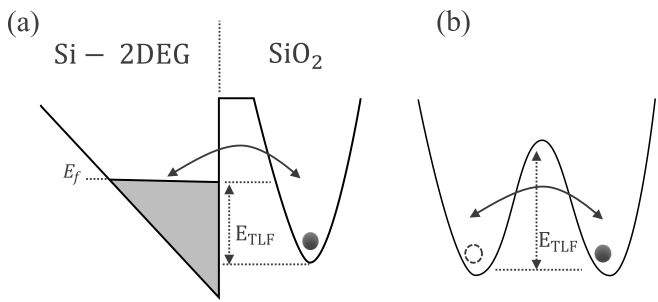}
\caption{Schematic models of two types Two-level Fluctuators (TLFs). (a) The TLFs can be the switching of the charge state of trap (whole charge fluctuation), (b) a single charge moving between the minima of double well potential (DWP) which can be found at the $\mathrm{Si/SiO_2}$ passivated interfaces \cite{biswas_hydrogen_1999}. TLFs have a characteristic activation energy, $\mathrm{E_{TLF}}$ which is overcome by thermal excitation or tunnelling under its corresponding energy barrier.}
\label{fig:TLFs}
\end{figure}

Structural defects, atomic vacancies and strain can modify the energy profile of a semiconductor and often act as charge traps \cite{fleetwood_effects_1993,fleetwood_defects_2009}. The capture and release of a charge from the a trap induces two-level electrostatic fluctuations essentially behaving as TLF. A TLF is usually identified by an activation energy $E_{TLF}$, which a charge needs to overcome (thermally or by tunnelling) to be captured in or released from the trap. TLFs are not limited to whole charge fluctuations in which a trap exchanges charges with an electron reservoir (2DEG) but it can also include dipolar fluctuations in which a single charge can alternate between two sites in double well potential (DWP) as shown in Fig.\ref{fig:TLFs}

As discussed earlier, we are interested in the electrostatic fluctuations caused by the TLFs which can be described by simple Coulombic potentials. The strength of an electrostatic fluctuation due to a whole charge trapping event as shown in Fig.\ref{fig:TLFs}(a) can be described by
\begin{equation}
    U_{WH}(\mathbf{r}) = \frac{e^2}{4\pi\epsilon_0\epsilon_r \mathbf{|r-r_t|}},
\label{eq_wholecharge}
\end{equation}
where $e$ is the electron charge, $\epsilon_0$ is the vacuum permittivity, $\epsilon_r$ is the relative permittivity of the host material and $\mathbf{r_t}$ is the position of the TLF. A charge alternating between two energy minima in a double well potential (DWP) as shown in Fig\ref{fig:TLFs}(b) causes electrostatic fluctuations described by 
\begin{equation}
    U_{DWP}(\mathbf{r}) = \frac{e^2}{4\pi\epsilon_0\epsilon_r  \mathbf{|r-r_t|}} - \frac{e^2}{4\pi\epsilon_0\epsilon_r \mathbf{|r-(r_t+r_l)|}},
\label{eq_DWP}
\end{equation}
where $\mathbf{r_t}$ is the position of one of the wells and $\mathbf{r_l}$ is the separation between the two wells among which the charge is alternating. A fluctuating dipole with both a positive and a negative charge would cause a electrostatic potential fluctuation two times the one of a single charge moving between two wells ($U_{dip} = 2$ $U_{DWP}$). 

Furthermore, a commonly used \cite{culcer_dephasing_2013,culcer_dephasing_2009,bermeister_charge_2014} description of the electrostatics of a TLF is  one of a whole charge trap (Fig.\ref{fig:TLFs}(a)) screened by the 2DEG in the reservoirs and in the Thomas-Fermi approximation is given by the formula 
\begin{equation}
    U_{WHS}(\mathbf{r}) = \frac{e^2}{4\pi\epsilon_0\epsilon_r ((x-X_t)^2+(y-Y_t)^2)^{3/2}} (\frac{1+q_{TF} d}{{q_{TF}}^2}),
\label{eq_wh_sc}
\end{equation}
where $q_{TF} = 2/ a_B $ is the Thomas-Fermi wave vector with $a_B \approx 3$ $\mathrm{nm}$; the effective Bohr radius in silicon and $d$ is the depth of the trap in the oxide \cite{davies_quantum_1997}. The expression in Eq.\ref{eq_wh_sc} is derived under an approximation assuming large $|\mathbf{r-r_t}|$. Hence, from the point of reference of a quantum dot sensing such fluctuations, a screened potential implies that such switching events are occurring further outside the perimeter of the dot ($>50$ $\mathrm{nm}$ from its center) for them to be screened by the 2DEG.

\section{Noise Simulations}
\label{sec:Simulations}

In the context of studying spin qubits in quantum dots, charge noise is often introduced as dephasing, tunnelling or detuning noise \cite{huang_spin_2018}. The noise signal is either described as $A/f$ or by a RTS (Eq.\ref{eq_RTS_AS},\ref{eq_RTS_S}) \cite{culcer_dephasing_2009}. The amplitude of a single RTS is usually calculated within the Thomas-Fermi screening approximation (Eq.\ref{eq_wh_sc}) under the assumption of depleted quantum dot region where the charge impurities are only active above the electron reservoirs which provide the screening \cite{culcer_dephasing_2009,culcer_dephasing_2013,bermeister_charge_2014}.

Charge sensing techniques have been widely adopted to characterize charge noise in quantum dot devices and spin qubit environments. In Ref.\cite{connors_charge-noise_2022}, a match is observed between the spectra extracted from qubit noise spectroscopy and from charge sensing methods confirming a common noise environment between the qubit and charge sensing quantum dots. This method uses quantum dots in the coulomb blockade regime as charge sensors, offering a more rapid noise characterization method as compared to qubit noise spectroscopy and also widely shows 1/f-like noise spectra. Reported noise spectral densities at 1Hz range from approximately 0.5 to 10 $\mu eV/\sqrt{Hz}$ \cite{stuyck_integrated_2020,zwerver_qubits_2021,freeman_comparison_2016,kim_low-disorder_2019,connors_charge-noise_2022}. 

In the more rapid charge sensing technique, the noise is measured as an electrostatic fluctuation of the dot's electrochemical potential which is extracted via current fluctuations. In a perturbation approach, the fluctuations in the electrochemical potential $\mu(N)$ of a quantum dot (with N electrons) due to TLFs can be approximated by the energy fluctuation of the Nth electron on the dot \cite{hanson_2007}. And is described by

\begin{equation}
    \Delta \mu(N) = \bra{\Psi_N} U_{TLF} \ket{\Psi_N},
\label{eq_deltamu}
\end{equation}
where $\Psi_N$ is the wave function of the Nth electron on the dot and $U_{TLF}$ is electrostatic potential of the TLF(Eq.\ref{eq_wholecharge}, \ref{eq_DWP} \& \ref{eq_wh_sc}). When used for charge/noise sensing, quantum dots are operated in the many electrons regime as Single Electron Transistors (SETs). The exact number of electrons on such a dot is typically in the tens, however, unknown. Hence, the N in Eq.\ref{eq_deltamu} is arbitrary, yet high enough to assume a non-interacting electron gas in the quantum dot \cite{giuliani_vignale_2005}. Furthermore, at large electron numbers, the parity of the number N becomes irrelevant as the shell effects disappear due to disorder and irregularities in the confining and background potentials \cite{Vorojtsov_2004}. To represent such a wave function we choose one with a uniform probability distribution over the quantum dot area \cite{Vorojtsov_2004}. The wave function $\Psi_N$ is described by the harmonic oscillator eigen-functions shown in Eq.\ref{eq:Hermite}. Here we assume a symmetric dot with $\lambda_x = \lambda_y = \lambda$. The wave function parameters and a comparison with the case of a single electron quantum dot is discussed in Appendix.\ref{app:SimParam}. It is important to note here that the symmetric quantum dot simulated here for charge noise sensing is not part of the device in Fig.\ref{fig:3D}. Albeit, a charge/noise sensing quantum dot would typically be located approximately 100 nm away, in the y-direction, from the DQD device.

To simulate the noise spectral density sensed by a quantum dot, a Si-MOS device is assumed where the electron is confined 2-3 nm below the $\mathrm{Si/SiO_2}$ interface as per the out-of-plane wave functions estimated from the model in Sec.\ref{app:DQD Model}. A random distribution (in both space and switching times) of TLFs is created over an area of 200 nm x 200 nm in the middle of which the device lies. The total spectral density as sensed by the dot can be described by 

\begin{equation}
    S_{N}(f) =\sqrt{\sum_{i}^{N_t}\frac{ \Delta \mu_i^2 \tau_i}{4 + (2\pi f)^2 \tau_i^2}},
\label{eq_Simulation}
\end{equation}
where $N_t$ is the total number of TLFs in the given area and can be deduced from their assumed density for any of the calculations and $\tau_i$ is the mean switching time of each of them. Experimental studies of oxide traps and TLFs have shown a wide range of switching rates that can vary from microseconds to seconds \cite{hofheinz_individual_2006}. Furthermore, at low temperatures, anomalous switching behavior of the TLFs was observed which can be attributed to the domination of tunnelling over thermally activated fluctuations \cite{reinisch_microscopic_2006}. Therefore, in this work $\tau_i$'s within a given range are randomly distributed among the TLFs, following a log-uniform distribution, and accordingly any given density would reflect the density of TLFs within the chosen range of switching times. 

\begin{figure}[h!]
\includegraphics{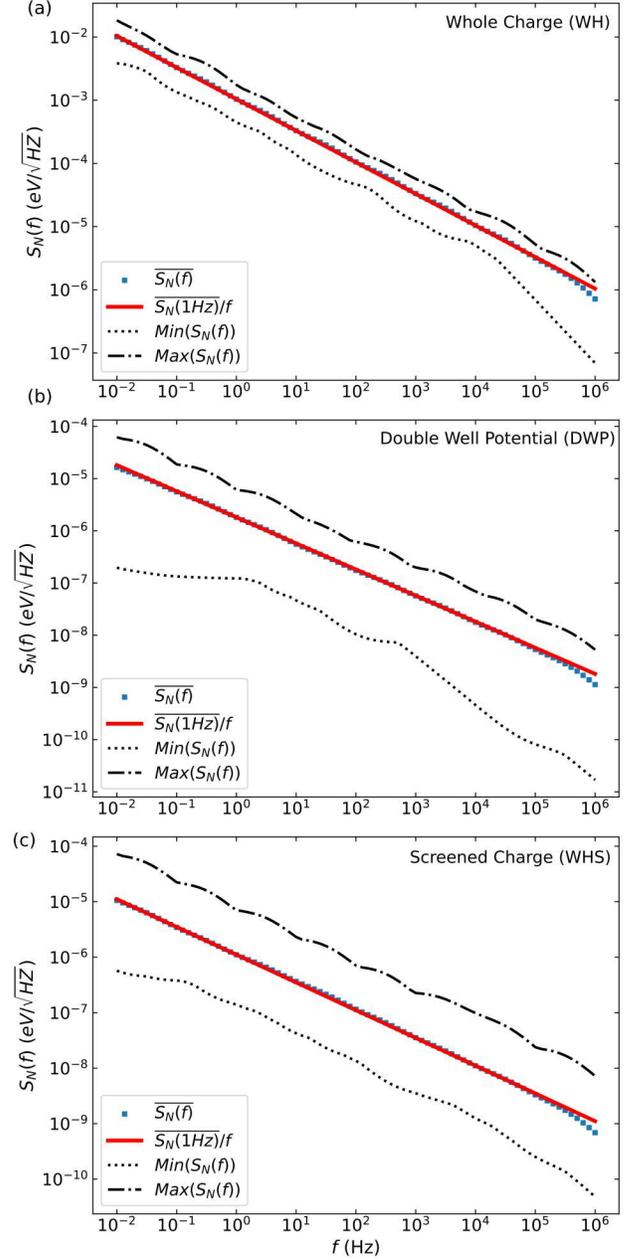}
\caption{Log-scale plot of the spectral densities calculated for 1000 random distributions of (a) un-screened whole charge fluctuations, (b) a charge tunnelling in a double well potential and (c) screened whole charge fluctuation within the Thomas-Fermi approximation. The regions between the dotted and dashed lines span the range from the minimum and maximum calculated values at each frequency, the blue squares mark the mean spectral density at every frequency and the red lines represent $A/\sqrt{f}$ where A is the mean spectral density at 1Hz. \label{fig:PSD_all} }
\end{figure}

\begin{figure}[h!]
\includegraphics{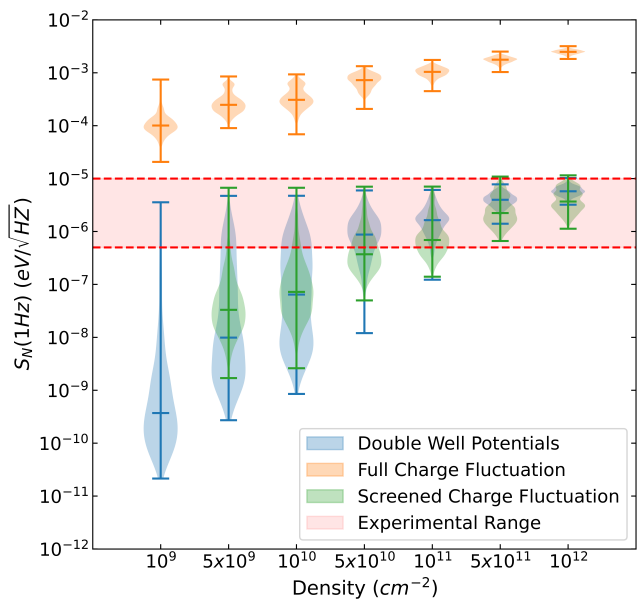}
\caption{Log-scale violin plot of the simulated noise spectral density at 1Hz vs the TLFs density for 1000 random TLF distributions per density. Each distribution is marked at its minimum, median and maximum values from bottom to top respectively. The shaded red region spans the experimentally reported range at the same frequency \cite{stuyck_integrated_2020,zwerver_qubits_2021,freeman_comparison_2016,kim_low-disorder_2019,connors_charge-noise_2022}. \label{fig:DensityVsNoise} }
\end{figure}

A Monte Carlo simulation is done with 1000 random distributions of each of the three different types of TLFs (un-screened whole charge, double well potential and screened whole charge) at a typical density of $10^{11}$ $\mathrm{cm^{-2}}$ \cite{culcer_dephasing_2013}. Each single simulation involves a number of TLFs (dictated by the density) with a random uniform distribution in space around the quantum dot and a randomly uniform distributed switching frequencies ($1/\tau$). The simulation results are shown in Fig.\ref{fig:PSD_all}. The regions between the dotted and dashed lines span the range between the minimum and maximum calculated values for the spectral density at a given frequency, the blue squares depict the mean spectral density at every frequency while the red lines are $A/\sqrt{f}$ guiding lines where $A$ is the mean value at $f=1Hz$. All three types of the TLFs are assumed to be at the $\mathrm{Si/SiO_2}$ interface and the DWPs are oriented along the out-of-plane (z) axis with a tunnel distance of 0.1 nm \cite{reinisch_microscopic_2006,biswas_hydrogen_1999,culcer_dephasing_2013}. Screened charge traps' distributions are restricted to the area at least $50$ $\mathrm{nm}$ away from center of the quantum dot to ensure the validity of the assumption of screening by the 2DEG \cite{culcer_dephasing_2009,culcer_dephasing_2013,bermeister_charge_2014}. The main figures of merit from the simulations in Fig.\ref{fig:PSD_all} are also shown in Table.\ref{tab:simulations}. It can be observed that the $A/\sqrt{f}$ line in Fig.\ref{fig:PSD_all} always lies within the range of simulated spectral densities at all frequencies, which is a confirmation of the validity of the distributed fluctuators model for representing 1/f-like charge noise in quantum dots. We repeat the procedure for different TLF densities and the distributions of the noise spectral densities at 1Hz are shown in Fig.\ref{fig:DensityVsNoise}

\begin{figure}[h!]
\includegraphics{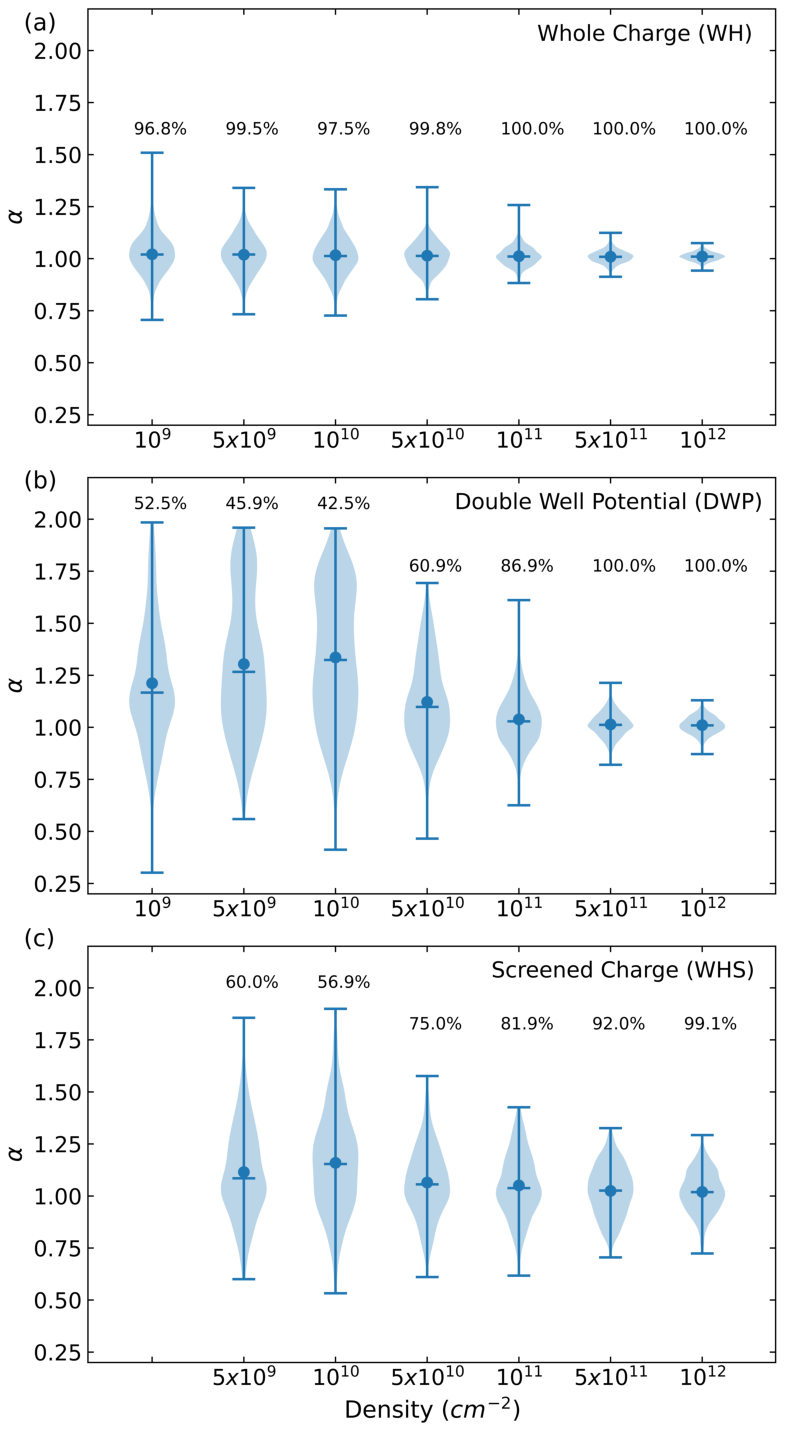}
\caption{Violin plots of the distributions of $\alpha$ extracted from a $A/f^{0.5\alpha}$ fitting of each of the 1000 individual noise spectral densities of (a) un-screened whole charge fluctuations, (b) a charge tunnelling in a double well potential and (c) screened whole charge fluctuation at different TLF densities. Each distribution is marked with a line at the minimum , median and maximum values and a dot at the mean value. The success rate of the linear fit is also shown above each distribution.\label{fig:Alpha_all} }
\end{figure}

The simulation results for whole charge fluctuations show spectral densities at 1Hz that are 2-3 orders of magnitude higher than the aforementioned experimentally reported values. Such a large difference is demonstrated at all TLF spatial densities. Accordingly, a charge noise model based on un-screened whole charge fluctuations can be dismissed given that large discrepancy between simulation and experiment. On the other hand, dipolar and screened charge fluctuations show spectral densities within a close range of the experimentally reported results for densities between $10^{11}$ and $10^{12}$ cm$^{-2}$. We further observe an expected larger variability of the noise metrics at low TLF densities which decreases as density increases. Such variability can be reflected on qubit devices performance and will be discussed in Sec.\ref{sec:Quantum Gate Fidelities}. It is worth noting that the total spectral density for each individual TLF distribution presented in Fig.\ref{fig:PSD_all} can potentially take any shape within the Min-Max range (see Fig.\ref{fig:RTNto1f}). Hence, larger ranges essentially indicate higher probability of deviations from the $A/\sqrt{f}$ trends. To demonstrate such deviations, each individual spectral density is fitted to an $A/f^{0.5\alpha }$ line using linear regression on a log scale and the values of $\alpha$ for fits with a coefficient of determination larger than or equal to 95\% are shown in Fig.\ref{fig:Alpha_all}. Each of the distributions in Fig.\ref{fig:Alpha_all} is denoted with the success rate of the linear regression fit. The values of $\alpha$ shown in the figure for whole charge fluctuations are relatively concentrated around the 1 which does not reflect the deviations from the standard $A/f$ spectra as reported in the literature \cite{connors_low-frequency_2019} unlike DWPs and screened charges which reflect a larger variation in $\alpha$. The mean value of $\alpha$ for the three types of TLFs is very close to 1 which further validates the charge noise mode presented here. Some spectra which didn't show a successful linear fit were randomly chosen and successfully fit to a Lorentzian function (Eq.\ref{eq_RTS_S}). This agrees with experimental observations of  RTS spectra\cite{kafanov_charge_2008,connors_low-frequency_2019,struck_low-frequency_2020}. 

\begin{table*}
\caption{\label{tab:simulations}
Noise spectral density simulations for whole charge, dipolar and screened charge TLFs at the given densities and switching frequencies. Reported here is the mean spectral density at 1 Hz ($\overline{S_{N}(1Hz)}$) in addition to minimum to maximum range at the same frequency for 1000 different distributions.}
\begin{ruledtabular}
\begin{tabular}{ccccc}
 &\multicolumn{2}{c}{Parameters}&\multicolumn{2}{c}{$S_N(f)$}\\
 TLF&Density($\mathrm{cm^{-2}}$)&Frequency Range(Hz)&$\overline{S_{N}(1Hz)}$
 &Min-Max at 1Hz\\ \hline
 Whole Charge&$10^{11}$&$10^{-2}-10^6$ & $\mathrm{1.04 meV/\sqrt{Hz}}$ & $\mathrm{0.45-1.75meV/\sqrt{Hz}}$\\
 DWP&$10^{11}$
&$10^{-2}-10^6$&$\mathrm{1.81\mu eV/\sqrt{Hz}}$ & $0.12-6.12 \mathrm{\mu eV/\sqrt{Hz}}$\\
 Screened Charge\footnote{Screened charges are assumed to be only above the 2DEG (Reservoirs) at least 50 nm to sides of the SET}&$10^{11}$&$10^{-2}-10^{6}$
 &$\mathrm{1.11\mu eV/\sqrt{Hz}}$&$\mathrm{0.14-7.06 \mu eV/\sqrt{Hz}}$\\
\end{tabular}
\end{ruledtabular}
\end{table*}

The simulations presented here offer a possible modeling framework for charge noise which matches results from experimental charge noise characterizations. As established from the simulations results, un-screened whole charge fluctuations (trapping and de-trapping) in the vicinity of a quantum dot induce electrostatic potential fluctuations that are 2-3 orders of magnitude higher than measured noise levels, a fact that can be easily checked by examining the values of $U_{WH}(r)$. Restricting such fluctuations to the area around the dots allows for screening by the 2DEG under the assumption of a quasi-depleted area in which a the quantum dots lie \cite{culcer_dephasing_2009,culcer_dephasing_2013,bermeister_charge_2014}. Furthermore, such spatial restriction allows for the use of the real space expression for the Thomas-Fermi approximation for screening (\ref{eq_wh_sc}) which is only valid for large $r$ ($q>2k_f$)\cite{davies_quantum_1997} and must be replaced with a momentum space representation at shorter distances. Despite the fact that distributions of screened charge fluctuations can simulate noise levels within the experimentally measured ranges, the spatial restrictions alongside the mathematical complications that come with them can prove impeditive to the inclusion of such potentials in a more elaborate quantum mechanical model of quantum dot devices as the one discussed in Sec.\ref{app:DQD Model}. Additionally, it would require the adjustment of the allowed regions for screened charges around quantum dot devices with different sizes, which implies a loss of generality in a noise model. For larger quantum dot networks, such as 2D arrays in large scaled systems, such an adjustment would position noise sources only at the edges of the array leaving the internal dots essentially noise free, which is a quite optimistic yet unrealistic assumption. Furthermore, the depleted dot assumption \cite{culcer_dephasing_2009,culcer_dephasing_2013,bermeister_charge_2014} does not necessarily hold for many electrons quantum dots such as single electron transistors (SETs) typically used for charge and noise sensing. Finally, simulated noise spectra for dipolar TLFs (DWPs) are within the experimentally measured ranges irrespective of any screening assumptions promoting it as versatile tool for modeling realistic charge noise that is easily integrable in quantum mechanical calculation of qubit systems as will be demonstrated in Sec.\ref{sec:Quantum Gate Fidelities}. Furthermore, a dipolar representation of charge noise sources can also describe screening of charge noise by metallic gates via image charges.

\section{Quantum Gate Fidelities}
\label{sec:Quantum Gate Fidelities}

A Universal set of quantum gates is one of the criteria for quantum computation as proposed by Divincenzo in 2000 \cite{divincenzo_physical_2000}. One and two- qubit gates can be used as the building blocks for a universal quantum computer fulfilling the aforementioned criterion. For instance, the one-qubit rotation gates ($X(\theta),Y(\theta),Z(\theta)$) alongside the two-qubit $\sqrt{SWAP}$, gate form a universal set of quantum gates \cite{williams2010explorations}. In this work we demonstrate the noise effect on the one-qubit X($\pi$) gate as an example of rotation gates in addition to the two-qubit SWAP gate whose half-operation would be the $\sqrt{SWAP}$. The SWAP gate is chosen given its nature as native exchange gate and can clearly demonstrate the effect noise on the exchange interaction. For a qubit encoded in a single electron spin, spin-to-charge coupling can be utilized for spin control. One-qubit gates can be implemented by EDSR spin control via intrinsic or artificially introduced SOI and two-qubit gates via the exchange interaction. Therefore, both gates' performance is expected to be limited by electrostatic fluctuations from TLFs which induce errors. 
Typically one- and two-qubit gates operate in MHz regime at which gate times would normally be less than a few microseconds\cite{veldhorst_addressable_2014,yoneda_quantum-dot_2018,zwerver_qubits_2021}. Accordingly, in the presence of low frequency fluctuators as in the simulations presented here, it is highly unlikely that more than a single switching event per TLF can occur within the gate time. Hence, such fluctuations would introduce measurement to measurement variations effectively decreasing quantum gate fidelities. In this study we estimate one- and two-qubit gate errors in the presence of dipolar TLFs (DWP) (Fig.\ref{fig:TLFs}(a)). We also study the effect of unscreened whole charge TLFs (WH) (Fig.\ref{fig:TLFs}(b)) in Appendix.\ref{Q_WH}. All quantum gates simulation are performed on the DQD device in Fig.\ref{fig:3D} with tunnel gate width of 20 nm. In Sec.\ref{sub_OneQubit}, we study a single quantum dot, with the DQD at a low coupling (high tunnel barrier) regime where an electron in each of the two dots can be operated independently as a single qubit. Single dot wave functions are calculated numerically and by diagonalizing the single electron Hamiltonian matrix in the same manner described in Sec.\ref{app:DQD Model}. In Sec.\ref{sub_TwoQubit}, we study two-qubit gates in the DQD system in a high coupling regime with exchange interactions and all wave functions parameters are calculated using the FCI model in Sec.\ref{app:DQD Model}. Noise is introduced in the one- and two- qubit systems in two manners. First, as individual TLFs to study the system's sensitivity to noise. Second, as random spatial distributions of TLFs with randomly distributed switching frequencies, which are introduced in time-dependent simulations to estimate quantum gate fidelities.
\subsection{One Qubit X-Gate}
\label{sub_OneQubit}
\begin{figure}[h!]
\includegraphics{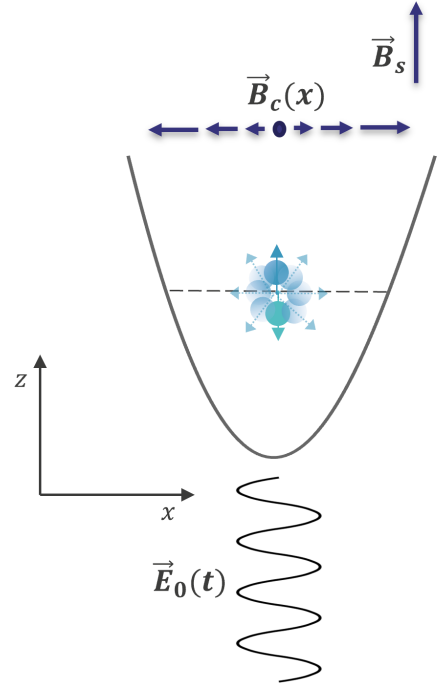}
\caption{Schematic illustration of EDSR-driven spin control in a quantum dot. $\mathbf{B_s}\hat{z}$ is the static Zeeman field, $\mathbf{B_c(x)}\hat{x}$ is the coupling magnetic field gradient and $\mathbf{E_0(t)}\hat{x}$ is the driving a.c. electric field.\label{fig:EDSR}}
\end{figure}

Using micro/nano magnets \cite{pioro_2008,Brunner_2011,Simion2020}, an electron in a quantum dot is subjected to a magnetic field gradient. Applying oscillating electric fields in resonance with the spin splitting energy (Zeeman Energy) induces an oscillation between the two spin states. The mechanism is Electric Dipole Spin Resonance (EDSR) \cite{rashba_theory_2008} illustrated in Fig.\ref{fig:EDSR} and can be described by the Hamiltonian 

\begin{equation}
\begin{split}
   \hat{H}_{EDSR}=\sum_{\zeta}^{x,y,z}&\frac{-\hbar^2}{2m_{\zeta}^*}\frac{\partial^2}{\partial\zeta^2} +V(\mathbf{r}) + g\mu_b \mathbf{B_s.S} \\& + g\mu_b \mathbf{B(r).S} +e\mathbf{E_0.r} cos(\omega t) +e\mathbf{E_n.r}.
\end{split}
\label{eq_EDSR_H}
\end{equation}
The first term in $\hat{H}_{EDSR}$ is the kinetic energy while $V(\mathbf{r})$ is electrostatic potential energy of an electron in the dot.  $V(\mathbf{r})$ is extracted from the DQD device in Fig.\ref{fig:3D} in a low coupling regime (high tunnel barrier/ low tunnel gate voltage) at which each of the two dots can be operated independently as single qubit.  At low energies, the eigen-functions of the first two terms in $\hat{H}_{EDSR}$ are obtained by solving the schrodinger equation on electrostatic potentials in Fig.\ref{fig:3D}. Such eigen-functions fit very well to those of the 2D harmonic oscillator in the x-y plane multiplied by $\xi(z)$ as shown in Fig.\ref{fig:OutOfPlane}. Hence, at low energies a separation of variables can be considered with $V(\mathbf{r})=V_{in}(x,y) + V_{out}(z)$ and the in-plane potential can be described as $V_{in}(x,y) = \frac{1}{2}m_{x(y)}^*(\omega_x^2x^2 + \omega_y^2 y^2)$. $m_{x(y)}^*=0.19m_0$ is the transverse electron mass and $\omega_{x,y}$ is the confinement frequency in the x and y directions respectively obtained by fitting the numeric eigen-functions to harmonic oscillator wave functions. An applied static magnetic field $\mathbf{B_s}$ provides a Zeeman splitting ($E_z = g\mu_B B_z$) between the two spin states where $g$ is the electron g-factor in silicon ($\approx$ $2$) and $\mu_b$ is the Bohr magneton. The magnetic field provided by the micro/nanomagnets, $\mathbf{B(r)}$, is spatially non uniform and $\mathbf{E_0}=(E_x,E_y,E_z)$ is the applied oscillating electric field with a frequency $\omega$. Finally $\mathbf{E_n}=(E_{n,x},E_{n,y},E_{n,z})$ is the electric field noise due to TLFs. When the resonance condition ($\omega =  g\mu_b |B_s|/\hbar $) is satisfied an X-Gate (NOT) operation can be achieved. 
The EDSR hamiltonian can be best described in the moving frame picture \cite{rashba_theory_2008} in which the wave functions are transformed as

\begin{equation}
\begin{split}
       & \Psi(r,t) \longrightarrow e^{-i \mathbf{k}.\mathbf{R}(t)}  \Psi(r,t) \\
       \mathrm{and} \\
       &r \longrightarrow e^{i \mathbf{k}.\mathbf{R}(t)}r e^{-i \mathbf{k}.\mathbf{R}(t)} = r+\mathbf{R}(t),\\
    \mathrm{where} \\
    &\mathbf{R}(t) = -e\mathbf{E_0} cos(\omega t) / m\omega_i^2 ,
\end{split}
\label{eq:transformation}
\end{equation}
with $\omega_i$ is the parabolic potential's confinement frequency in the direction of the driving field $\mathbf{E_0}$.
As described in Ref.\cite{rashba_theory_2008} the transformation above introduces two terms in the Hamiltonian. The first term $-e^2E_0^2cos^2(\omega t)/2m\omega_0^2$ which only introduces global energy shift and has no effect in spin-gate operations. The second term $-2\frac{e\hbar\omega}{m\omega_0^2}(\mathbf{k.E_0})sin(\omega t)$ can be neglected as $\omega/\omega_0 \ll 1$ in this study. In this work the driving electric field is chosen to be in the x-direction giving $\mathbf{E_0}=(E_x,0,0)$ and $\mathbf{R}(t)=X(t)=-eE_x cos(\omega t)/m\omega_x^2$. 

The Hamiltonian matrix in the basis of the spin states $\ket{\uparrow},\ket{\downarrow}$ in the dot's ground state can be described as 
\begin{equation}
\frac{g\mu_b}{2}
\begin{pmatrix}
B_z & B_c(X(t))\\
B_c(X(t)) & -B_z
\end{pmatrix}.
\label{eq:EDSR_matrix}
\end{equation}

The external static magnetic field is chosen to be out-of-plane (z-direction); $\mathbf{B_s}=(0,0,B_z)$ \cite{Simion2020}. Hence, the spin-coupling magnetic field  $\mathbf{B_c}(i) = \sqrt{(\partial_i B_x i)^2+(\partial_i B_y i)^2} \ni i\in \{x,y,z\}$ and the dephasing magnetic field, $\mathbf{B_{deph}}= \mathbf{\nabla} B_z.\mathbf{r}$. The static magnetic field ($\mathbf{B_s}=B_z$) is chosen to be 0.3T and the magnetic field gradients are chosen from micro-magnetic simulations \cite{dumoulin_stuyck_low_2021} as

\begin{equation}
\begin{pmatrix}
\partial_xB_x & \partial_xB_y & \partial_xB_z\\
\partial_yB_x & \partial_yB_y & \partial_yB_z\\
\partial_zB_x & \partial_zB_y & \partial_zB_z\\
\end{pmatrix}
=
\begin{pmatrix}
0.5&0.05&0.05\\
0.05&0.5&0.05\\
0.05&0.05&-1\\
\end{pmatrix}
\mathrm{mT/nm}.
\label{eq:Field_Gradient}
\end{equation}

An external driving field $E_x = \mathrm{7.5KV/m}$ gives a an EDSR displacement amplitude $X(t)=X(0) \approx 0.5$ $\mathrm{nm}$ leading to a Rabi frequency of 3.85 MHz and a spin flip time $t_X \approx 0.13 \mu s$. Note that here we only include the coupling field gradients in the direction of the driving electric field ($B_c(x)$).

In realistic conditions where $E_n\neq0$, the effective noise on the qubit can be either transverse or longitudinal to its quantization axis \cite{ramon_2022}. Hence, the charge noise can induce gate errors in two ways. First, via modulating the driving in-plane electric field (Transverse Noise) and hence the Rabi frequency of the spin oscillation which dictates the quantum gate time. Second, via modulating the out-of-plane electric field (Longitudinal Noise) which induces fluctuations in the Zeeman splitting in the presence of a dephasing magnetic field gradient ($\mathbf{\nabla}B_z$) which in turn causes fluctuations in the resonance frequency. It is important to note here that we assume that the DQD system stays in a low coupling regime in the presence of noise. Accordingly, each dot stays in a single-qubit operation mode for which we perform the single-qubit gate simulations. In this section we define our origin (0,0,0) at the center of the single dot of interest for the reader to easily visualize the position of a noise source relative to the single qubit. For the rest of the paper, the origin is defined at the center of the DQD system as shown in Fig.\ref{fig:3D}.

\subsubsection{Coupling (Transverse) noise}
To study the spin-coupling (transverse) noise, the noise term $E_n$ is introduced in the qubit matrix (Eq.\ref{eq:EDSR_matrix}) by projecting the Hamiltonian in Eq.\ref{eq_EDSR_H}  on the single qubit basis using a Schrieffer Wolf transformation changing $X(t)\longrightarrow X'(t) = X(t) + d_{n,x}(t)$  with $d_{n,x}(t) = \frac{e}{m\omega_x} \int_{0}^{t} d\tau sin(\omega_x\tau)\langle E_{n,x}(t-\tau)\rangle$. The electric field in the direction of drive ($E_{n,x}$) due to a TLF is assumed constant over a single quantum dot area via integrating it over its ground state wave function. Unlike the ideal system in Eq.\ref{eq:EDSR_matrix}, the electric field from TLFs can displace the electron in all three dimensions ($\mathbf{d_n}=(d_{n,x},d_{n,y},d_{n,z})$), hence, the coupling magnetic fields $B_c(y) \& B_c(z)$ would introduce further transverse noise. However, for simplicity, only the transverse noise due to displacement in the direction of the driving electric field is included in this work.

Electrostatic simulations of  a single quantum dot  from the DQD device in Fig.\ref{fig:3D} give the lowest single electron eigen-functions of a single dot with $\omega_x/2\pi \approx \mathrm{0.56THz}$ and $\omega_y/2\pi \approx \mathrm{2.4THz}$. The charge noise induced errors on an EDSR driven X-gate are evaluated by allowing the qubit to evolve in time using the time dependent Schrodinger equation with the Hamiltonian matrix in Eq.\ref{eq:EDSR_matrix} in the presence of a single TLF. The qubit state at $t_X$ is then compared to noise-free time evolved state and the error is calculated using the formula

\begin{equation}
\label{eq:Error}
Error = 1-\frac{|\Psi_N(t_{X})|^2}{|\Psi_I(t_{X})|^2},
\end{equation}
where $\Psi_I$ and $\Psi_N$ are the single electron spin states of the ideal and noisy system respectively.
Table.\ref{tab:EDSR} shows the error induced on a X-gate in the presence of single TLF. Due to the resonance condition imposed by the Zeeman splitting between the two spin states, the results of our simulations show that a single qubit gate can show relatively low sensitivity to low frequency coupling charge noise with errors $\ll 1\%$. While highly unlikely, a TLF with a switching frequency in resonance with the qubit would alter the spin flip frequency by a factor $\approx \langle E_{n,x}\rangle/E_{0,x} $ which would have a deleterious effect from a TLF in close vicinity of the qubit. 

However, in the case of a TLF resonant with the qubit, tuning the Zeeman splitting via the static magnetic field can bring the system to a more stable condition. Furthermore, given that $d_{n,i} \propto 1/\omega_i^2$, the qubit insensitivity to spin coupling noise can be further enhanced by stronger confinement in the quantum dots. For example, the devices simulated here (Fig.\ref{fig:3D}) have an asymmetric confinement with $(\omega_y/\omega_x)^2 \approx 20$.  Consequently, an EDSR drive in the y-direction (($\mathbf{E_0}=(0,E_y,0)$) would render the qubit frequency less sensitive to charge noise (Demonstrated in App.\ref{Q_WH}). However, in such case the amplitude of the driving electric field $E_y$ would need to be 20 times larger to achieve the same EDSR displacement amplitude ($X(0)$) which can be experimentally challenging due to cross couplings in a device's gate space.

\begin{table}[h]
\caption{\label{tab:EDSR}%
Coupling Noise induced X-Gate errors is the presence of one TLF}
\begin{ruledtabular}
\begin{tabular}{ccddd}
TLF&(x,y,z)\footnote{The single quantum dot is centered at (0,0,0) and is 2 nm below the $\mathrm{Si/SiO_2}$ interface} nm&
\multicolumn{1}{c}{\textrm{Gate Error \%}}\\
\hline
DWP&(10,0,2)&0.001 \\
\end{tabular}
\end{ruledtabular}
\end{table}
\subsubsection{Dephasing (Longitudinal) noise}
\begin{figure}
    \centering
    \includegraphics{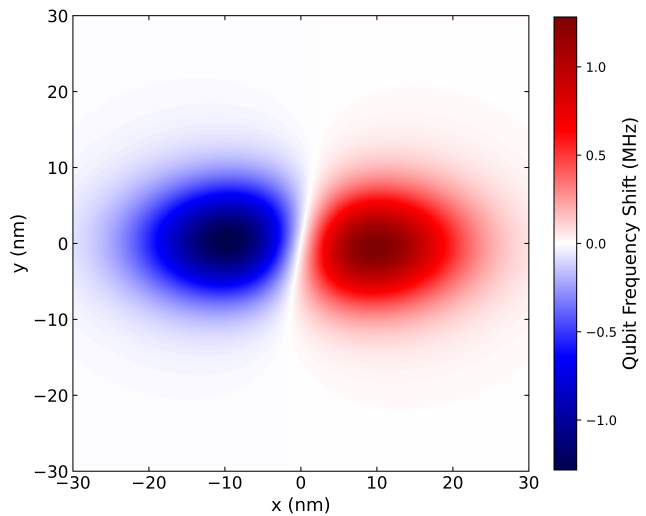}
    \caption{Qubit frequency shift as function of the position of a DWP indicating sensitivity to the quantum dot confinement. Larger sensitivity is demonstrated to x-polarized electric fields, the direction of least confinement for the devices simulated here. A 2D electron wave function located 2 nm below the $\mathrm{Si/SiO_2}$ interface is assumed for the quantum dot.}
    \label{fig:FShifts_2D}
\end{figure}

As discussed earlier in the text, the electrostatic fluctuations due to the TLFs can displace the electron in three dimensions with a displacement amplitude vector $\mathbf{d_n}$ whose components for longitudinal noise are given by
\begin{equation}
    d_{n,i}=\frac{e \langle E_{n,i}\rangle}{m\omega_i^2} \ni i \in \{x,y,z\},
\label{eq:displacement}
\end{equation}
which leads to a shift in the qubit's resonance frequency $\Delta\omega=g\mu_b(\mathbf{d_n.B_{deph}})/2\hbar$. It is important to note that $\omega_z$ is included in Eq.\ref{eq:displacement} just to give a complete description while in reality the out-of-plane (z-direction) confinement is not parabolic. In Si-MOS devices, it is best described by triangular potentials due to the step at the $\mathrm{Si/SiO_2}$ interface in addition to the applied gate voltages and Si/SiGe implementations would be described by tilted quantum wells. Furthermore, given that the electron confinement is strongest in the z-direction, longitudinal noise is expected to be dominated by in-plane electric fields from the TLFs. Here we calculate $d_{n,(x,y)}$ using Eq.\ref{eq:displacement} with the electric field due to the TLFs, $E_{n,(x,y)}$, assumed constant by integrating over the ground state wave function of the dot and  $\omega_x$ and $\omega_y$ extracted numerically from the device's potential as described earlier in this section. The out-of-plane eigensystem is solved numerically and the displacement $d_{n,z}$ is estimated in the presence of the TLFs' out of plane field ($E_{n,z}$) which is found, as predicted, small compared to its in-plane counterparts, however, shall not be ignored given that the dephasing gradient $\partial_zB_z$ is 20 times the in-plane dephasing gradients (Eq.\ref{eq:Field_Gradient}). Table.\ref{tab:EDSR_deph} shows the resonance frequency shifts and gate errors induced by a single TLF's dephasing electric field. Each of the TLFs is positioned to have a strongly polarized electric field to study the correlation between the displacement direction and X-gate fidelity. Similar to the coupling noise case, given the $1/\omega_i$ dependence of the displacement of the electron due to noise $d_{n,i}$, it is expected that sensitivity of the qubit frequency would increase as the confinement of the quantum dot decreases. For the simulated devices with asymmetric confinement, the simulation results show that the qubit is most sensitive to electric fields polarized in the dot-separation direction (x-direction) due to the relatively weak confinement in such direction in comparison to the y- and z-directions. As shown in Table.\ref{tab:EDSR_deph}, a DWP with  x-polarized electric field causes a frequency shift and a gate error percentage around 5 times and 30 times its y-dominant counterpart, respectively. Such polarization dependence is further demonstrated in Fig.\ref{fig:FShifts_2D} with a 2D approximation of the electron's wave function where the qubit frequency shift is visualized as a function of the DWP position at the $\mathrm{Si/SiO_2}$ interface. It can be seen that a DWP directly above the dot at the origin doesn't cause in-plane displacement of the electron and hence the qubit frequency is effectively unchanged.

\begin{table}[h!]
\caption{\label{tab:EDSR_deph}%
Dephasing noise induced qubit Frequency shifts and X-Gate errors is the presence of one TLF}
\begin{ruledtabular}
\begin{tabular}{cccc}
TLF&(x,y,z)\footnote{The single quantum dot is centered at (0,0,0) and is 2 nm below the $\mathrm{Si/SiO_2}$ interface} nm& $\Delta f (Hz)$&
\multicolumn{1}{c}{\textrm{Gate Error \%}}\\
\hline
DWP&(0,0,2)&1.97*$10^5$&0.27\\
DWP&(0,5,2)&1.75*$10^5$&0.21\\
DWP&(5,0,2)&9.37*$10^5$&5.8 \\
\end{tabular}
\end{ruledtabular}
\end{table}

To obtain a more experimentally relevant evaluation of the single qubit gate operation, a random distribution of TLFs is created in the manner described to Sec.\ref{sec:Simulations} and with randomly distributed switching frequencies in the range assumed in Table.\ref{tab:simulations} ($10^{-2} - 10^{6}$ $\mathrm{Hz}$). The quantum gate is simulated using the \textit{QuTip} package in \textit{python}. The qubit is initialized in one of the spin states and is allowed to evolve for the calculated X-gate time $t_X$ for 1000 consecutive gate operations in the presence of the fluctuating fields from the TLFs. The qubit fidelity in the given randomly generated noise environment is given by 
\begin{equation}
\label{eq:Fidelity_S}
F = \frac{|\Psi_N(t_X)|^2}{|\Psi_I(t_X)|^2}*100,
\end{equation}
where $\Psi_I$ is the single electron state of the ideal and $\Psi_N$ is state in the noisy system averaged over 1000 X-gate operations. For each kind of TLF, two random distributions are generated and the X-gate fidelity is calculated twice per distribution with two different random allocations of switching frequencies. Table.\ref{tab:EDSR_Fid} shows X-gate fidelities due to DWPs with a spatial density of $10^{11}$ $\mathrm{cm^{-2}}$. The X-gate fidelity in Table.\ref{tab:EDSR_Fid} is calculated  with two different random distributions of interfacial DWPs (1 and 2) and two different random distributions of switching frequencies (a and b) per spatial distribution. The number of distributions here is limited to four given the computational time cost of the time dependent simulations.

\begin{table}[h]
\caption{\label{tab:EDSR_Fid}%
Single Qubit X-Gate fidelities in presence of switching TLFs distribution with a density of $\mathrm{10^{11} cm^{-2}}$}
\begin{ruledtabular}
\begin{tabular}{cccc}
TLF&Spatial Dist. &Switching Freq. Dist.& Fidelity\%\\
\hline
\multirow{4}{*}{DWP}&\multirow{2}{*}{1}&a&99.999\\
&&b&99.51\\
&\multirow{2}{*}{2}&a&97.55\\
&&b&99.1\\
\end{tabular}
\end{ruledtabular}
\end{table}

\begin{figure}[h]
\includegraphics{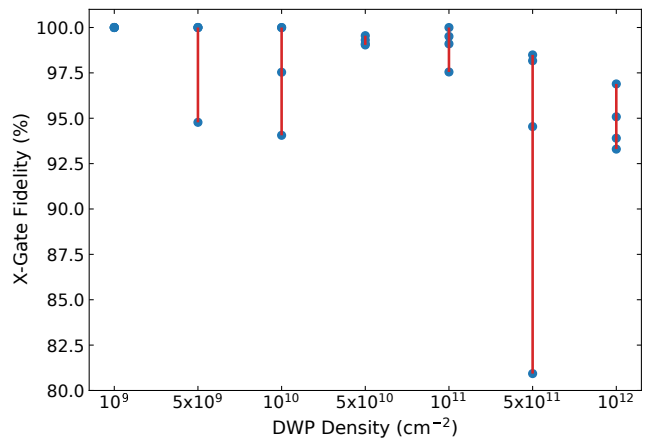}
\caption{Average X-Gate fidelities for the simulated distributions at different DWP densities \label{fig:X-GateF_Den} }
\end{figure}

We further investigate the X-gate fidelity at different spatial densities of DWPs and the results are shown in Fig.\ref{fig:X-GateF_Den} and Appendix.\ref{sec:FidelityVsDensity}. Similar to the simulations presented in Table.\ref{tab:EDSR_Fid}, four different simulations are performed per spatial density. It is important to note that device variability can be quite significant especially at lower densities \cite{wu_variability_2020} due to the $1/r^2$ dependence of the electric field noise which makes the quantum gate fidelity highly dependent on the state and switching frequencies of its closest TLFs. For instance, a device with only one trap close to it that is active during the simulation time (1000 X-gate operations) might have a lower gate fidelity than a device with tens of TLFs surrounding which are mostly further from it or inactive during the simulation time. Given the time cost of such simulations it was impractical to adopt a statistical approach by simulating thousands of devices as in Ref.\cite{wu_variability_2020}. However, the figures we present here give a clear indication of the performance variability even at low TLF densities. Nevertheless, despite the statistical variability, one might still expect an average statistical correlation between noise spectral densities and quantum gate fidelities. In App.\ref{app:FidelityVsSpectra} we adopt a slightly simplified approach to calculate the X-gate fidelities and study the correlation between the spectral densities of electrostatic (and qubit frequency) noise and X-gate fidelity.

Electron confinement energies in qubit architectures are limited by fabrication technologies in addition to design requirements. For instance, for a single qubit in a DQD device as the one simulated here, it is expected that the electrons are to be less confined in the direction in which the quantum dots are separated and/or the direction in which their interaction shall be mediated, a feature that can be more pronounced in the aspired larger quantum dot arrays for scalable quantum computation. Micro/nano-magnet designs that minimize the dephasing fields in the dot-separation direction can effectively reduce the single-qubit sensitivity to charge noise. For example, in Ref.\cite{dumoulin_stuyck_low_2021}, a cobalt micro-magnet design is proposed to minimize the dephasing fields in the y-direction to almost zero due to the pseudo-encapsulation of the DQD by the micro-magnet in that direction which leads to cancellation of the dephasing fields. If rotated by $90^{\circ}$, such a micro-magnet can show better noise resistance and can sustain single qubit gate fidelities greater than 99\% irrespective of the displacement directions.
It is worth noting that the error mechanisms discussed here still hold for Electron Spin Resonance (ESR) implementations of single spin control in which effective magnetic field gradients can come from nuclear fields or spin-orbit coupling.

\subsection{Two Qubit SWAP-Gate}
\label{sub_TwoQubit}

Exchange driven SWAP gates for spin qubits were proposed by Divincenzo and Loss in 1998 \cite{loss_quantum_1998}. The interaction between the spins can encapsulated in the Heisenberg Hamiltonian 

\begin{equation}
    \hat{H_s}= J(t) \mathbf{S_1.S_2},
\label{eq_Heisnberg}
\end{equation}
and the two qubit gate matrix described in the $\ket{\uparrow\downarrow'}$ \& $\ket{\downarrow\uparrow'}$ basis can be described as 

\begin{equation}
\frac{1}{2}\begin{pmatrix}
0 & J(t)+\Delta J(t)\\
J(t)+\Delta J(t) & 0
\end{pmatrix},
\label{eq:Exchange_Matrix}
\end{equation}
where  $\ket{\uparrow\downarrow'} = (\ket{S_0}+\ket{T_0})/\sqrt{2}$, $\ket{\uparrow\downarrow'} = (\ket{S_0}-\ket{T_0})/\sqrt{2}$, $J(t)$ is the exchange interaction and $\Delta J(t)$ is the shift due to charge noise. To demonstrate the effect of noise on exchange, we choose the native exchange SWAP gate. Hence, we assume that the Zeeman splitting is the same in both dots ($\Delta E_z = 0$). In gate defined quantum dots, $J(t)$ can be controlled via tunnel gate voltages or via the energy detuning ($\epsilon_d$) between the two dots and in a first order approximation $J \propto t^2/(U-\epsilon_d)$ \cite{huang_spin_2018}. Pulsing the exchange coupling for a time t such that $ \int_{0}^{t} J(t')/\hbar \,dt'  = \pi (\mathrm{mod}$ $2\pi)$, exchanges the spin states of the two electrons demonstrating a SWAP gate.
Exchange interaction, $J$, for a double quantum dot architecture has been calculated, under symmetric operation, using the developed full configuration interaction model (Sec.\ref{app:DQD Model},Fig.\ref{fig:Exchanges}). For a DQD device (Fig.\ref{fig:3D}) with 20 nm tunnel gate width, tuning the tunnel gate voltage to 0.97V gives a tunnel barrier of approximately 3 meV leading to a calculated exchange interaction $J \approx 18.28$ MHz which can drive a SWAP operation in 27 $ns$. Electrostatic fluctuations can alter the tunnel coupling (tunnel noise) or the detuning (detuning noise) between the two dots \cite{huang_spin_2018}. 

In this work, the effect of TLFs on two-qubit gates is evaluated in two ways. First, exchange interaction is calculated using the aforementioned quantum mechanical model of the DQD system after introducing the full 3D electrostatic potential of a single TLF. The results of such calculations are compared to the exchange interaction in the ideal system and are shown in Table\ref{tab:Exchange}. Calculations done in the presence of a DWP right above the center of the DQD system (tunnel noise dominant) and above one of the dots (detuning noise dominant) showing a stronger effect of tunnelling noise with a 15\% shift in exchange interaction as compared to a 8.5\% shift with a detuning noise as theorized in \cite{huang_spin_2018}. We further compare the effect of a TLFs electric field direction by positioning a DWP 50 nm away from the one of the quantum dots in x- and y-directions which showed a stronger shift in exchange (0.0195\%) for an x-positioned DWP as compared to a y-positioned one (0.00113\%). As for the case of single qubit, the devices simulated here (Fig.\ref{fig:3D}) have a stronger confinement in the y-direction as compared to the x-direction which shows that exchange's sensitivity to noise is also has an inverse correlation to the quantum dots' confinement. However, in both cases the shift in exchange is rather small as compared to tunnelling and detuning noises. 

Second, a random distribution (in space and switching frequencies) of DWPs is created in a similar manner as in Sec.\ref{sec:Simulations} and the system is allowed to evolve in time for the quantum gate time dictated by the exchange interaction in the ideal system ($t_{SWAP}= 27 ns$) in the presence of time dependent alternating electrostatic potential of the TLFs. The two-qubit state is averaged over 1000 two-qubit operations and the gate fidelities shown in Table.\ref{tab:ExchangeF} are calculated in comparison to the ideal gate operation with the formula
\begin{equation}
\label{eq:Fidelity}
F = \frac{|\Psi_N(t_{SWAP})|^2}{|\Psi_I(t_{SWAP})|^2}*100,
\end{equation}
where $\Psi_I$ and $\Psi_N$ are the two electron states of the ideal and noisy system respectively.  It is worth mentioning here that unlike the one-qubit gate simulations, in two-qubit gate simulations, noise is introduced in the DQD system with full 3D potentials of the noise sources under no assumption of a constant potential over the DQD region. 

Due to the low frequency nature of the TLFs introduced to the system, the closest TLFs to the DQD system have the dominant effect on the two-qubit gate fidelity due to the strong measurement to measurement variation they can introduce as per the results in Table \ref{tab:Exchange}. Intuitively, when operating in the symmetric regime (tunnel control of exchange), higher J could lead to a smaller modulation ($\Delta J/J$) due to charge noise. However, exchange calculations shown in Fig.\ref{fig:Exchanges} show that tunability of the exchange interaction is limited by the hybridization of two quantum dots into a single dot indicated by an onset of saturation in its values. Furthermore, the zero energy splitting between the two-qubit gate states imposes no resonance condition on the SWAP gate operation and hence the high sensitivity to charge noise. Other two-qubit gate implementations such as the CZ and CPHASE gates \cite{veldhorst_two-qubit_2015,petit_high-fidelity_2020}, rely on a the difference of Zeeman energy ($\Delta E_z$) between the two dots. In this case the charge noise enters the system via two channels: the exchange (J) as described and demonstrated here for the SWAP gates and $\Delta E_z$ as the electron is displaced in the magnetic field gradient. An interplay between the two mechanisms is expected here and a detailed study of CZ and CPHASE fidelities is left for future works.

\begin{table}[h]
\caption{\label{tab:Exchange}%
Shift in the exchange interaction calculated using a full configuration interaction model in the presence of TLFs in close vicinity of the DQD system}
\begin{ruledtabular}
\begin{tabular}{ccddd}
TLF&(x,y,z) nm\footnote{The DQD is centered at (0,0,0) and is 2 nm below the $\mathrm{Si/SiO_2}$ interface}&
\multicolumn{1}{c}{\textrm{$\Delta J/J$}}\\
\hline
DWP&(0,0,2)& 0.15 \\
DWP&(30,0,2)& 0.085 \\
DWP&(80,0,2)&1.95*10^{-4}\\
DWP&(30,50,2)&1.13*10^{-5}\\
\end{tabular}
\end{ruledtabular}
\end{table}

\begin{figure}[h]
\includegraphics{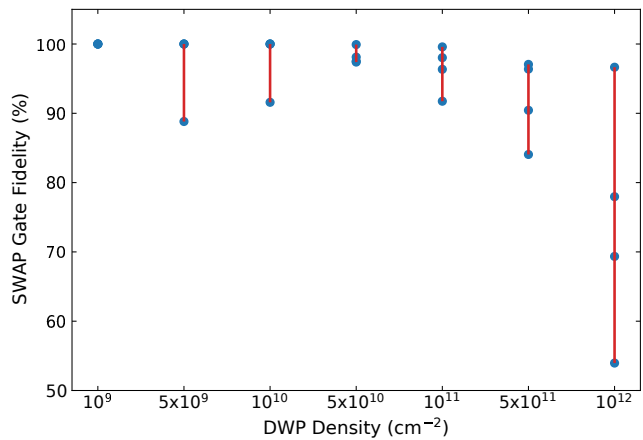}
\caption{Average SWAP Gate fidelities for the simulated distributions at different DWP densities \label{fig:SWAP-GateF_Den} }
\end{figure}

We finally study the SWAP gate fidelity at different DWP densities and the results are shown in Fig.\ref{fig:SWAP-GateF_Den} and Appendix.\ref{sec:FidelityVsDensity} indicating a similar conclusion to the results of one-qubit gate simulations. The $1/r$ dependence of the TLFs' Coulombic potential is a key metric to charge noise in quantim dots. For the Si-MOS devices simulated here, electrostatic fluctuations from TLFs can drop a two-qubit SWAP gate fidelity below 90\% even at TLF densities as low as $5*10^9$ $\mathrm{cm^{-2}}$ which can be challenging to practically achieve. To summarize, the results presented here provide a quantifiable demonstration of the intuitive conclusion that reduction of disorder in devices in addition to moving the physical qubit device from the noise sources are key improvements towards higher qubit device yields. 


\begin{table}[h]
\caption{\label{tab:ExchangeF}%
SWAP Gate fidelity in the presence of TLFs distrubutions}
\begin{ruledtabular}
\begin{tabular}{ccddd}
TLF&Density ($\mathrm{cm^{-2}}$)&
\multicolumn{1}{c}{\textrm{Fidelity \%}}\\
\hline
DWP&$10^{11}$&91.7\\
\end{tabular}
\end{ruledtabular}
\end{table}

\section{Conclusions}
\label{sec:Conclusions}
To summarize, we offer here a co-modeling framework for spin qubits in semiconductors and their noise background. The modeling framework is demonstrated in this work on Si-MOS quantum dots and can be extended to other quantum dot systems such as in Si/SiGe and GaAs/AlGaAs. Electrostatic potentials of realistic device architectures (Fig.\ref{fig:3D}) can be extracted from electrostatic device simulations. The single electron solutions of such potentials are used to model single qubits. Furthermore, we develop a versatile full configuration interaction quantum mechanical model of double quantum dot (DQD) devices which takes the numerical electrostatic potentials as an input and estimates exchange interaction efficiently. The DQD model further allows the inclusion of noise potentials to directly study the effect of individual noise sources on the device operation parameters. Based on experimentally reported noise spectra, we represent charge noise based on a distributed fluctuations model. The results of electrostatic noise simulations suggest that dipolar TLFs (DWPs) offer a close representation of charge noise at low temperature especially when compared to the unrealistically strong fluctuations caused by un-screened whole charge traps in addition to the restrictive conditions implied by screening assumptions.  In an EDSR implementation of single qubit control, and an exchange controlled two qubit interaction, we find an inverse correlation between the quantum dot confinement frequencies and their sensitivity to charge noise. The resonance condition imposed by the Zeeman splitting renders one-qubit X-gates relatively insensitive to coupling noise. On the other hand, dephasing noise can cause shifts in the qubit resonance frequency, which, as mentioned above, is more pronounced as the quantum dot confinement frequency is decreased.  The absence of a resonance condition in addition to the electrostatic nature of the exchange interaction puts a SWAP gate under the mercy of charge noise which is naturally dominated by the closest TLFs to the DQD system. We also study the one- and two-qubit gates' fidelities in the presence of different randomly distributed TLFs' densities. We find that given the small size of quantum dots, the quantum gate performance is dominated by the TLFs closest to the quantum dot, which results in large variability in the qubit performance. For instance, as shown in Tables \ref{tab:XGatevsDensity} and \ref{tab:SWAPGatevsDensity}, quantum gate fidelities can vary by more than $10\%$ at the same TLF density. Such results indicate that qubit device architectures that physically position the quantum dots as far as possible from sources of noise can offer a more robust implementations of scalable spin in quantum dot-based quantum computations. 

Upon the completion of this manuscript, the authors have come across the work in Ref.\cite{buonacorsi_optimizing_2020} in which as similar "central basis" approach had been independently developed and adopted to model double quantum dot devices

\begin{acknowledgments}
We thank Karam Gamaleldin for his help with the illustrations. We also thank Alexander Grill for the insightful discussions. This work was supported in part by the imec Industrial Affiliation Program on Quantum Computing.
\end{acknowledgments}

\newpage
\appendix
\begin{widetext}
\section{Effective Mass Approximation and Valley Interference}
\label{app:EMA}

As mentioned in the main text, the DQD model presented here solves the two-electron Hamiltonian within the effective mass approximation. In other words, the electron states are described using envelope wave functions. In reality the full electron wave function in silicon has the following form 

\begin{equation}
\label{eq:fullWF}
\psi_{\mu}(\mathbf{r}) = \mathbf{u_{k_{\mu}}(r)} e^{i\mathbf{k_{\mu}.\mathbf{r}}}F_{\mu}(\mathbf{r}),
\end{equation}
where $\mathbf{u_{k_{\mu}}(r)} e^{i\mathbf{k_{\mu}.r}}$ are the Bloch wave functions which are solutions of the silicon bulk crystal Hamiltonian at any of the 6 degenerate valleys denoted by $\mu$ ($\mu  \in $\{$\pm x, \pm y , \pm z$ \}). $F_{\mu}(\mathbf{r})$ are the envelope wave functions computed within the effective mass approximation and are shown in Fig.\ref{fig:OutOfPlane} and Fig.\ref{fig:InPlane} of the main text. The 6 conduction band minima (valleys) are located at $\mathbf{k_{\mu}} \approx 0.85 a_0 \hat{\mu}$ which introduces a periodicity in the electron wave function that is non-commensurate with the lattice periodicity. Therefore, valley interference effects may appear \cite{Koiller_2001,Calderon_2008}, leading to oscillations in the overlap of wave functions with respect to the distance between them. Such oscillations can be reflected in relevant parameters such as tunnel couplings and in turn, exchange. In gate defined quantum dots, the out-of-plane confinement alongside the effective mass anisotropy lift the 6-fold degeneracy leaving the $\mathrm{\pm z}$ valleys as the lower energy valleys with $\approx$ 30 meV energy separation from the other 4 valleys. With that, one can expect that interference effects would be manifested in inter-valley interactions over the out-of-plane (z-) direction. In the case of laterally coupled quantum dots where the interactions are mediated over the in-plane (xy-) direction, we expect that valley interference effects to be exponentially suppressed by the $\mathrm{e^{ik_zz}}$ term. Below we demonstrate this by calculating all the integrals involved in the DQD model in Sec.\ref{app:DQD Model} using the full wave function in Eq.\ref{eq:fullWF}.

We start by studying the valley coupling due to the single electron terms of the Hamiltonian in Eq.\ref{eq:twoelectronH}. Under the separation of variables assumption,

\begin{equation}
V(\mathbf{r})=V_z(z)+V_{x}(x)+V_{y}(y),
\end{equation}

and

\begin{equation}
F_{\mu}(\mathbf{r})=X_{\mu}(x)Y_{\mu}(y)Z_{\mu}(z)
\end{equation}

where $\mathrm{V_{\mu} (\mu)}$ is the electrostatic potential in the $\hat{\mathrm{\mu}}$ direction. First we compute the valley coupling due to the out-of-plane potential given by 

\begin{equation}
VC_{z}= \bra{\psi_{+z}(\mathbf{r})} V_z(z) \ket{\psi_{-z}(\mathbf{r})} = \sum_{G,G'} c^{*}_{+z}(\mathbf{G}) c_{-z}(\mathbf{G^{'}}) I_x I_y I_z
\end{equation}

Here we used the plane wave expansion of the Bloch wave functions as described in Ref.\cite{SaraivaValley} by replacing  $\mathbf{u_k(r)} $ with $ \sum_G c_{\mu}(\mathbf{G}) e^{i\mathbf{G.r}}$, where $\mathbf{G}= (G_x,G_y,G_z)$ are reciprocal lattice vectors. Therefore ,
\begin{equation}
\label{eq:IX}
I_{x} = \int dx X^*_{+z}(x)X_{-z}(x) e^{i(G^{'}_x-G_x)x}.
\end{equation}

Using the expansions described in App.\ref{app:Coeff}, the integral above becomes simply the Fourier transform of a Gauss-Hermite function which is also a Gauss-Hermite function in reciprocal space. For quantum dots which are larger than a few a Bohr radii,($\lambda > a_B$)  $\mathrm{I_x}$ is effectively zero except when $\mathrm{G^{'}_x =G_x}$ and $\mathrm{X_{-z} = X_{+z}}$ where it is equal to 1 simply satisfying the normalization condition. The same argument applies for $\mathrm{I_y}$, therefore,

\begin{equation}
I_x I_y I_z= \delta(G^{'}_x-G_x)\delta(G^{'}_y-G_y)I_z.
\end{equation}

The equation above is identical to one described in Ref.\cite{SaraivaValley}, where 

\begin{equation}
I_{z} = \int dz Z^{*}_{+z}(z) Z_{-z}(z) e^{i(G^{'}_z-G_z-2k_z)z} V_z(z).
\end{equation}

$V_z$ here is simply a step potential due to the oxide interface in addition to linear potential due to the electric field from gates as described in Ref.\cite{SaraivaValley}. The sharp potential step at the $\mathrm{Si/SiO_2}$ interface (or due to band offset in Si/SiGe quantum wells) introduces a coupling between the two valley states. Under the assumption that $\mathrm{Z_{-z}(z) = Z_{+z}(z)}$, which is expected given the large energy spacing the z-direction, we calculate the valley coupling to be $\mathrm{VC_{z}\approx 65 e^{i \phi_z} \mu eV}$ where $\phi_z$ is the valley coupling phase and is  $\approx 0.62\pi$.

Second, in a similar manner we calculate the valley coupling due to the in-plane potentials given by
\begin{equation}
\label{eq:VCXY}
VC_{xy}=VC_{x}+VC_{y}= \bra{\psi_{+z}(\mathbf{r})} V_{x}(x)+V_{y}(y) \ket{\psi_{-z}(\mathbf{r})} = \sum_{G,G'} c^{*}_{+z}(\mathbf{G}) c_{-z}(\mathbf{G^{'}}) (J^x_x J^x_y J^x_z+J^y_x J^y_y J^y_z)
\end{equation}

To allow analytic analysis, here we use parabolic potentials as described in Ref.\cite{White_2018} instead of the numerical potentials to describe the DQD, which we believe will give the same qualitative results in the context of valley interference as long as the numerical potentials don't have any in-plane sharp features. With that we can describe the potentials as follows 

\begin{equation}
\begin{split}
&V_x(x)=\frac{m^{*}_x \omega^2_x}{2} \frac{1}{4d^2} (x^2-d^2)^2, \\
&\mathrm{and}\\
&V_y(y)=\frac{m^{*}_y \omega^2_y}{2}y^2,\\
\end{split}
\end{equation}

where d is the distance between the two dots. Accordingly,

\begin{equation}
J^{x}_{x} = \frac{m^{*}_x \omega^2_x}{8 d^2} \int dx X^*_{+z}(x)X_{-z}(x) e^{i(G^{'}_x-G_x)x} (x^2-d^2)^2,
\end{equation}

which is analytically solvable and with $\lambda > a_B $ the integral is effectively zero except when $G^{'}_x-G_x=0$ where the integral is simply the matrix element between the two envelope wave functions. The same result applies for $\mathrm{J^y_y}$, while $\mathrm{J^x_y}$ and $\mathrm{J^y_x}$ are equal to $\mathrm{I_x}$ and $\mathrm{I_y}$ respectively and $\mathrm{J^x_z =J^y_z=J_z}$ simplifying Eq.\ref{eq:VCXY} to 

\begin{equation}
\label{eq:VCXY2}
VC_{x}+VC_{y} =\sum_{G,G'} c^{*}_{+z}(\mathbf{G}) c_{-z}(\mathbf{G^{'}}) \delta(G^{'}_x-G_x)\delta(G^{'}_y-G_y) ( \bra{X_{+z}(x)} V_{x}(x) \ket{X_{-z}(x)} +\bra{Y_{+z}(y)} V_{y}(y) \ket{Y_{-z}(y)}) J_z,
\end{equation}

with 

\begin{equation}
J_z=\int dz Z^{*}_{+z}(z) Z_{-z}(z) e^{i(G^{'}_z-G_z-2k_z)z}.
\end{equation}

Given the strong confinement in the z-direction, $\mathrm{J_z}$ is not necessarily exponentially suppressed at non-zero powers of the complex exponential. However, by numerically calculating this integral we find it is always less than $10^{-2}$. Calculating the full summation in Eq.\ref{eq:VCXY2}, we find it further suppressed to be effectively zero. 

We that we conclude that valley coupling due the single electron Hamiltonian is no more than the coupling due to the Si/SiO2 interface which leads to two valley superposition states 

\begin{equation}
\psi_{\pm}(\mathbf{r},\phi_z)= \frac{1}{\sqrt{2}}(\psi_{+z} (\mathbf{r})e^{i \phi_z} \pm \psi_{-z}(\mathbf{r})),
\label{eq:valleysuperposition}
\end{equation}

that are eigen-functions of the full single electron Hamiltonian of the crystal in addition to the externally induced potentials at the interface. The two valley-split states are energetically separated by valley splitting which is $\mathrm{=2VC_z}$. 

Next we study the valley coupling due to the Coulombic interaction. Deviations from the single-valley integrals in Sec.\ref{app:DQD Model} which are calculated within the effective mass approximation can be introduced via the valley-flip terms 

\begin{equation}
\label{eq:vf}
\begin{split}
 &v_{1f}=\bra{\psi_{\pm z}(r_1)\psi_{ \pm z}(r_2)} \frac{e^2}{4\pi\epsilon r} \ket{\psi_{\pm z}(r_1)\psi_{\mp z}(r_2)},\\
 &v_{2f}=\bra{\psi_{\pm z}(r_1)\psi_{ \pm z}(r_2)} \frac{e^2}{4\pi\epsilon r} \ket{\psi_{\mp z}(r_1)\psi_{\mp z}(r_2)},
\end{split}
\end{equation}

in addition to the valley exchange terms

\begin{equation}
\label{eq:ve}
v_{e}=\bra{\psi_{\pm z}(r_1)\psi_{ \mp z}(r_2)} \frac{e^2}{4\pi\epsilon r} \ket{\psi_{\mp z}(r_1)\psi_{\pm z}(r_2)}.
\end{equation}

Expanding the in-plane envelope wave functions as described in Sec.\ref{app:DQD Model} and App.\ref{app:Coeff}, Coulombic integrals in Eq.\ref{eq:vf},\ref{eq:ve} take the following form 

\begin{equation}
\begin{split}
&\frac{e^2}{4\pi\epsilon} \sum_{n,m,l,p} \mathcal{C}\sum_{G,G',K,K'} c^{*}_{\mu_1}(\mathbf{G}) c_{\mu_2}(\mathbf{G^{'}}) c^{*}_{\mu_3}(\mathbf{K}) c_{\mu_4}(\mathbf{K^{'}})\\ 
& \int dXdYdxdydz_1dz_2 \frac{\psi_l(X,Y) e^{iS_xX+S_yY}\psi_p(x,y)e^{iD_xx+D_yy} |\xi(z_1)|^2 |\xi(z_2)|^2 e^{i\mathcal{G} z_1} e^{i\mathcal{K} z_2}}{\sqrt{x^2+y^2+(z_1-z_2)^2}}\\
&\mathrm{with}\\
&S_i= G^{'}_i-G_i+K^{'}_i-K_i  \ni i \in \{x,y\},\\
&D_i= G^{'}_i-G_i-K^{'}_i+K_i  \ni i \in \{x,y\},\\
&\mathcal{G}=G^{'}_z-G_z-sign(\mu_1)(1-\delta_{\mu_1,\mu_2})2k_z\\
&\mathcal{K}=K^{'}_z-K_z-sign(\mu_3)(1-\delta_{\mu_3,\mu_4})2k_z
\end{split}
\label{eq:CoulV}
\end{equation}

where $\mathrm{X=\frac{x_1+x_2}{2}}$ and $\mathrm{x=x_1-x_2}$ (similarly for y) . All the Hermite polynomial expansions are absorbed in the summation, $\sum_{n,m} \mathcal{C}$ for the sake of space and are explained in more details in App.\ref{app:Coeff}. The integration over X and Y resemble $\mathrm{I_x}$ in Eq.\ref{eq:IX} and are zero except when $\mathrm{S_x=S_y=0}$. Accordingly, this leaves $\mathrm{D_i = 2(G^{'}_i - G_i) = 2(K_i-K^{'}_i)}$ and Eq.\ref{eq:CoulV} becomes 

\begin{equation}
\begin{split}
&\frac{e^2}{4\pi\epsilon} \sum_{l,p} \mathcal{C}\sum_{G,G',K,K'} c^{*}_{\mu_1}(\mathbf{G}) c_{\mu_2}(\mathbf{G^{'}}) c^{*}_{\mu_3}(\mathbf{K}) c_{\mu_4}(\mathbf{K^{'}}) \mathcal{F} \{ \psi_l(X,Y) \}(0,0) \\ 
& \int dxdydz_1dz_2 \frac{\psi_p(x,y) |\xi(z_1)|^2 |\xi(z_2)|^2 e^{i\mathcal{G} z_1} e^{i\mathcal{K} z_2}}{\sqrt{x^2+y^2+(z_1-z_2)^2}},\\
\end{split}
\label{eq:CoulV2}
\end{equation}

where $ \mathcal{F}$ is the Fourier transform. Since the two relevant valleys here are in the z-direction, valley interference effects may appear due to the complex exponentials in $\mathrm{z_1}$ and  $\mathrm{z_2}$ . Hence, to maximize such effect we set  $\mathrm{D_x}$ and  $\mathrm{D_y}$ to zero and numerically calculate the integral in Eq.\ref{eq:CoulV2} at different values of $\mathrm{r^2=x^2+y^2}$ where x=y for simplicity.  Fig.\ref{fig:ValleyInteraction} shows the results of the calculations as compared to a same valley calculation $v_s$,  showing that $v_{2f}$ and $v_e$ are effectively zero and can be ignored. Single valley flip terms $v_{1f}$ are approximately two orders of magnitude lower than same valley ones, and are expected to be even lower in an integration over the three dimensions where $D_x$ and $D_y$ are not equal to zero.

To conclude, given the calculations presented here and in Ref.\cite{jiang_coulomb_2013,culcer_2010,hada_eto_2005,Koiller_2001,tariq_2022}, we believe that the effective mass approximation is valid for laterally coupled quantum dots with interfaces that can lift the 6-fold valley degeneracy. Accordingly, envelope wave functions can suffice to quantum mechanically study such systems.

\begin{figure}[h!]
\includegraphics{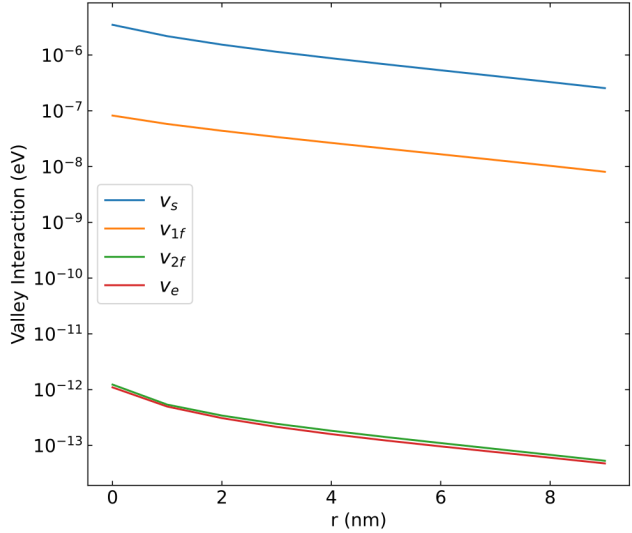}
\caption{A log-scale plot of $v_{1f}$, $v_{2f}$ and $v_{e}$ from Eq.\ref{eq:vf},\ref{eq:ve} vs the in-plane separation $r$. When compared with the same valley Coulombic interaction ($v_s$) , valley flip and valley exchange terms are much smaller and can be neglected. \label{fig:ValleyInteraction} }
\end{figure}

\section{Exchange With Steps at $\mathrm{\mathbf{Si/SiO_2}}$ Interface - Simplified model}
\label{app:Valley_Exchange_Steps}
As earlier described in the main text the exchange interaction (J) can be described as $J \propto t^2/U$ \cite{huang_spin_2018} at zero energy detuning between the dots, where $t$ is the tunnel coupling and $U = U_s-U_d$ is the difference between the Coulombic energy of two electron on the same dot ($U_s$) and different dots ($U_d$) respectively. This expression is a first order approximation of the expression that has the form \cite{veldhorst_two-qubit_2015}
\begin{equation}
J= \frac{1}{2}  [U - \sqrt{16t^2 + U^2}],
\label{eq:Hubbard_Exchange}
\end{equation}
which can be easily calculated by diagonalizing the Hubbard Hamiltonian.
In Ref.\cite{tariq_2022}, the effect of the difference of valley phase between the two dots is studied and it is shown that a difference of $\mathrm{\pi}$ would suppress the tunnel coupling and hence the exchange interaction. This can be visualized in Fig.\ref{fig:Valley_Tunnel} where the single-valley tunnel coupling ($\mathrm{t_0}$) defines the energy separation between the symmetric and anti-symmetric solutions of the the double quantum dot system as shown in Fig.\ref{fig:InPlane}. In a multi-valley system, namely two-valleys as in the case of the devices studies here, a difference in valley-phase and/or valley splitting between the two dots can lead to a tunnel coupling between the lower valley-split states at the two dots, $\mathrm{t}$, that is not equal to $\mathrm{t_0}$ and can even be suppressed down to zero as discussed in Ref.\cite{tariq_2022}.

\begin{figure}[h]
\includegraphics[scale=1]{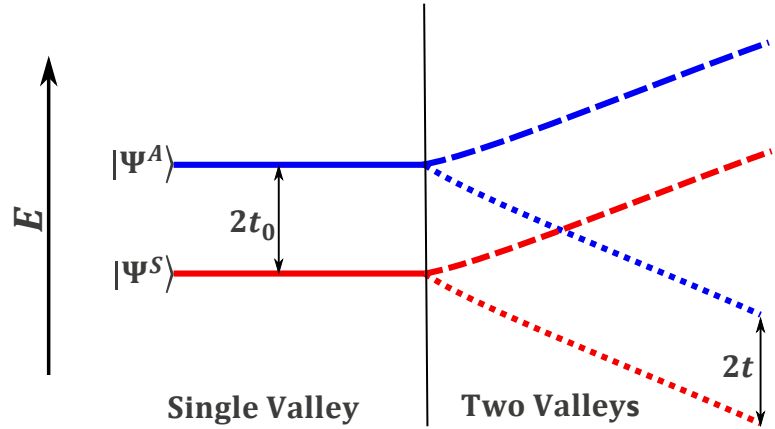}
\caption{A schematic energy diagram describing the tunnel coupling in a single-valley regime ($\mathrm{t_0}$,left) and in the two-valley regime ($\mathrm{t}$,right)\label{fig:Valley_Tunnel}}
\end{figure}

Here we use a simplified analytic approach to study the effect of an atomic step at the $\mathrm{Si/SiO_2}$ interface between the two dots. This approach is first order perturbative and relies on the assumption that the atomic step at the interface is at the tail regions of the quantum dot wave functions and does not include inter-orbital interaction between valleys. Hence, the effect would be proportional the position of the step relative to the overlap region of the two quantum dots wave functions. The difference in valley splitting between the symmetric and anti-symmetric  ($\Delta_S-\Delta_A$) and the difference in valley phase ($\phi_S-\phi_A$) are then calculated.  Shown in Fig.\ref{fig:Atomic_Step} is a schematic of the two quantum dots lying below the interface with an atomic step defined between them. The step is always assumed to be between the dots (i.e. the step is not right on top of any of the two quantum dots to maintain validity of the approach employed here). A step too close to the center of a quantum dot would drastically alter the potential profile and the device viability.

\begin{figure}[h]
\includegraphics[scale=1]{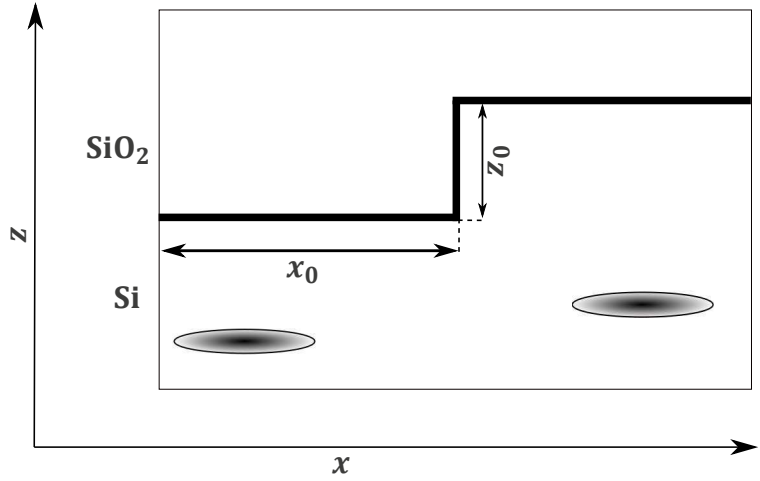}
\caption{A schematic diagram (not to-scale) of two quantum dots defined under a $\mathrm{Si/SiO_2}$ in a MOS device. A step at the interface with a depth $\mathrm{z_0}$ is introduced to study its effect on the valley structure of the two dots. The step is defined between the two dots at $\mathrm{x_0}$. The zero point on the x-axis ($\mathrm{x_0=0}$) is defined when the step is exactly in the middle between the two dots.)\label{fig:Atomic_Step}}
\end{figure}

The valley coupling is then calculated as
\begin{equation}
\Delta_{S(A)}=\bra{\Psi^{S(A)}_{+z}} V(z) \ket{\Psi^{S(A)}_{-z}},
\label{eq:ValleyC_SA}
\end{equation}
where $\mathrm{V(z)}$ is the out-of-plane potential including the step potential at the interface in addition to the atomic step with depth $\mathrm{z_0}$ positioned at  $\mathrm{x_0}$ and the wave functions $\mathrm{\Psi_{S(A)}}$ are defined as 

\begin{equation}
\Psi^{S(A)}_{\pm z} =  \mathbf{u_k(r)} e^{i\mathbf{\pm k_{z}.\mathbf{r}}} \frac{1}{\sqrt{2}\sqrt{1+(-)\eta}}(\psi^L +(-) \psi^R),
\label{eq:ValleyCouplingStep}
\end{equation}

where  $\mathrm{\psi^{L(R)}}$ is the ground state wave function of the left (right) quantum dot and $\mathrm{\eta}$ is the overlap between the two wave functions ($\mathrm{\bra{\psi^L}\ket{\psi^R}}$). $\mathrm{V(z)}$ is assumed here to take the form 
\begin{equation}
V(z) = U_0 [\theta(z)\theta(x_0-x)+\theta(z-z_0)\theta(x-x_0)],
\end{equation}

where $\mathrm{U_0}$ is the size of the potential barrier at the interface (3 eV) and $\theta$ is the unit step function. Using a Hubbard-like approach using the two lowest single electron eigen functions of the DQD ($\mathrm{\Psi^S \& \Psi^A}$) at the lowest valley split state ($\psi_-$ in Eq.\ref{eq:valleysuperposition}), the two electron energy spectrum is calculated by diagonalizing the two electron Hamiltonian matrix in the zero detuning regime ($\mathrm{\epsilon_d=0}$) and under the assumption that the self-energy ($U_s$) is the same for both dots. The two electron Hamiltonian in the matrix form is

\begin{equation}
\begin{pmatrix}
\frac{U_s+U_d}{2}-2t & 0 & 0& \frac{U_s-U_d}{2}F_1(\phi_S,\phi_A)\\
0 & \frac{U_s+U_d}{2} &\frac{U_s-U_d}{2}F_2(\phi_S,\phi_A)& 0\\
0&\frac{U_s-U_d}{2}F_2(\phi_S,\phi_A) &\frac{U_s+U_d}{2} &0\\
\frac{U_s-U_d}{2}F_1(\phi_S,\phi_A) &0 & 0&\frac{U_s+U_d}{2}+2t\\
\end{pmatrix}
\begin{pmatrix}
\ket{\psi^S_- \uparrow,\psi^S_- \downarrow }\\
\ket{\psi^S_- \uparrow,\psi^A_- \downarrow }\\
\ket{\psi^S_- \downarrow,\psi^A_- \uparrow }\\
\ket{\psi^A_- \uparrow,\psi^A_- \downarrow }\\
\end{pmatrix},
\label{eq:Field_Gradient}
\end{equation}

with 

\begin{equation}
\begin{split}
&F_1(\phi_S,\phi_A) =\frac{1}{4}( e^{i (\phi_S-\phi_A})+1)^2 ,\\
&F_2(\phi_S,\phi_A) =\frac{1}{4}(e^{-i (\phi_S-\phi_A)}(e^{-i (\phi_S-\phi_A)}+1)^2 .
\end{split}
\end{equation}
$\mathrm{F_{1,2}}$ are calculated by approximating all the highly oscillatory terms to zero given the absence of steps or high frequency components in the two-electron Coulombic interaction. The exchange is then calculated to be 
\begin{equation}
J= \frac{1}{2} [U cos(\phi_S-\phi_A)^2 - \sqrt{16t^2+U^2cos(\phi_S-\phi_A)^4}],
\label{eq:Exchange_Hubbard_Valley}
\end{equation}
which represents the exchange in the same form as Eq.\ref{eq:Hubbard_Exchange} introducing the effect of the difference in valley phase between the symmetric and anti-symmetric wave functions.
Fig.\ref{fig:ValleyPhaseDifference} shows the calculated difference in magnitude and phase of the valley coupling (Eq.\ref{eq:ValleyCouplingStep}) between the symmetric and anti-symmetric wave functions at an inter-dot distance of 25 nm and 50 nm. It can be observed that at the inter-dot distance of 25 nm, the difference in valley coupling is an order of magnitude lower than typical tunnel couplings which would be in the hundreds of $\mu eV$. The valley phase difference is observably much smaller than $\pi$ except in the region at $\mathrm{z_0 \approx 0.8 nm }$ and $\mathrm{-15 nm < x_0 < 0 nm}$ where the valley phase difference appears to be close to $\pi$. In this region the valley coupling difference is almost zero and we attribute the large difference in valley phase to the failure of the perturbative approach utilized here and would be resolved by introducing higher order wave functions and their interactions. Furthermore, it is important to note that an inter-dot distance of 25 nm is unrealistically small and the region at which the assumption of having the atomic step only affecting the wave function's overlap region is limited. On the other hand, at an inter-dot distance of 50 nm (which is still less than the smallest distance in the devices simulated in the model demonstration in Sec.\ref{app:DQD Model}), both the difference in valley coupling and valley phase due to the atomic step are negligible. With that, we can deduce that an effective mass approach is still valid even in the presence of an atomic step between the dots as long as the distance between the dots is not smaller than 50 nm. It is worth noting here that in realistic devices the interface disorder is not limited to atomic steps and to the region between the dots. However, a disorder that is strong enough to alter the valley spectrum enough to suppress exchange interaction, is likely also strong enough to be detrimental for proper quantum dot formation and a qubit demonstration in the first place. Moreover, the modeling framework presented here aims at studying the operation of DQD devices as a function of physical and operational parameters in addition to its sensitivity to charge noise. Studying such devices in the presence of interface disorder and at the limit of valley interference due to structural non-idealities is left for future works.

\begin{figure}[h]
\includegraphics[scale=1]{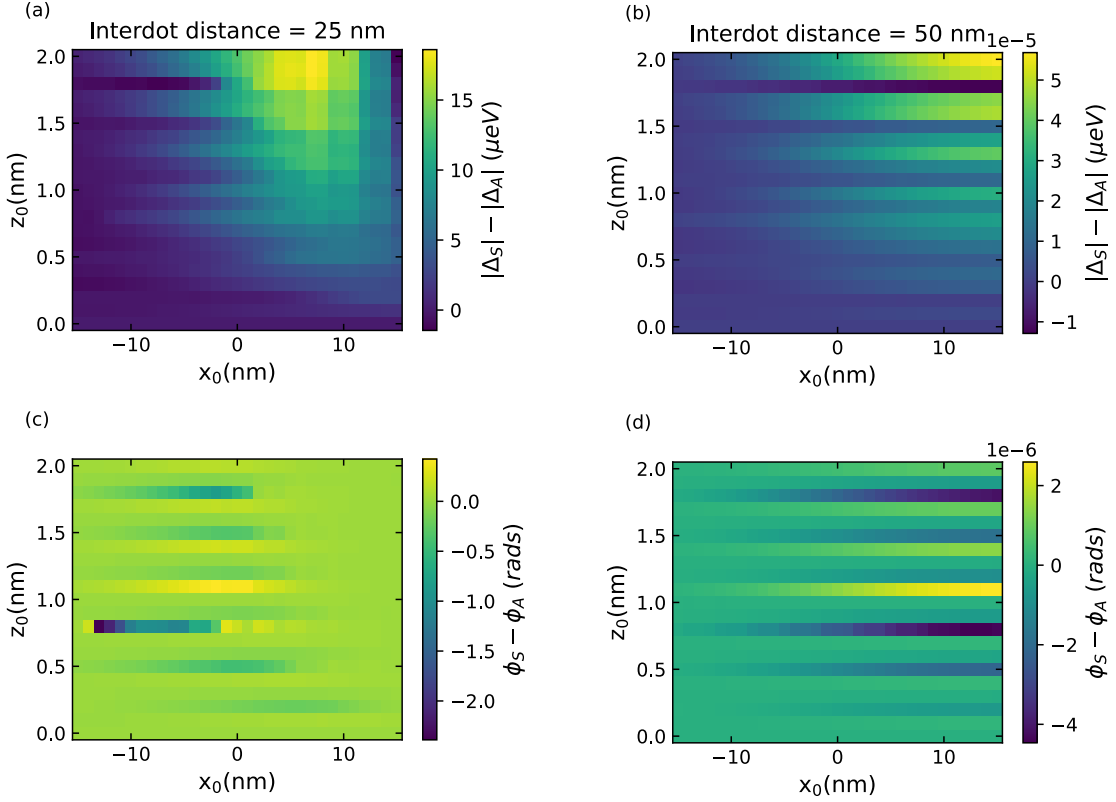}
\caption{2D color maps of the difference of valley coupling (a,b) and valley phase (c,d) between the symmetric and anti-symmetric double quantum dot states ($\mathrm{\Psi^S \& \Psi^A}$) calculated at an inter-dot distance of 25 nm (a,c) and 50 nm (b,d) for different depths of the atomic steps ($\mathrm{z_0} $ and at different positions ($\mathrm{x_0}$)). \label{fig:ValleyPhaseDifference}}
\end{figure}
\newpage

\section{Hermite Polynomials Expansion}
\label{app:Coeff}
Coulombic integrals in the basis described in Eq.\ref{eq_singlewavefunction} and Eq.\ref{eq:Hermite} have the form 

\begin{equation}
\begin{split}
\int &dx_1\,dx_2\,dy_1\,dy_2\,dz_1\,dz_2 \\ &\psi_{n_1}^{*}(x_1,y_1)\xi_{n_1^{z}}^{*}(z_1)\psi_{n_2}(x_1,y_1)\xi_{n_2^{z}}(z_1)\frac{e^2}{4\pi\epsilon\sqrt{(x_1-x_2)^2+(y_1-y_2)^2+(z_1-z_2)^2}}\psi^{*}_{m_1}(x_2,y_2)\xi_{m_1^{z}}^{*}(z_2)\psi_{m_2}(x_2,y_2)\xi_{m_2^{z}}(z_2),
\end{split}
\label{eq:CoulCart}
\end{equation}
which is the same integral in Eq.\ref{eq:Coul} described Cartesian coordinates, with $n_i = (n_{i}^x,n_{i}^y)$ and similarly  $m_i = (m_{i}^x,m_{i}^y)$. Given that all wave functions used here are real, complex conjugation will be dropped for the rest of the derivation. Since $\psi(x,y)$ can be rewritten as $\psi(x)\psi(y)$ we will describe the expansion only in $x$ which is identical to the one in $y$. 

Using the generating functions of the Hermite polynomials described by 

\begin{equation}
e^{2xt-t^2}=\sum_n \frac{H_n(x)}{n!} t^n,
\label{eq:generator}
\end{equation}

the product $\psi_{n_1}(x_i)\psi_{n_2}(x_i)$ can be expanded as

\begin{equation}
\sum_n^{n1+n2} C_n \psi_n(x_i),
\label{eq:expansion1D}
\end{equation}
with  
\begin{equation}
\begin{split}
 C_n = &\sqrt{\frac{n_1!n_2!n!}{\lambda\sqrt{2\pi}}}(\frac{1}{\sqrt{2}})^{n_1+n_2}\\ &\sum_{p_1,p_2}^{n_1+n_2} 2^{\frac{-p1-p2}{2}} \frac{1}{p_1!} \frac{1}{p_2!}\frac{1}{(\frac{n+n_1-n_2-p_1+p_2)}{2}) !} \frac{1}{(\frac{n-n_1+n_2+p_1-p_2)}{2}) !}\frac{1}{(\frac{-n+n_1+n_2-p_1-p_2)}{2}) !} H_{p_1}(0)H_{p_2}(0) 
 \end{split},
\end{equation}
and 
\begin{equation}
    \psi_n(x_i) = \frac{1}{\mathcal{N}} e^{\frac{-x_{i}^2}{2 \lambda_x^{'2}}}H_n(\frac{x_i}{\lambda_x^{'}}) ; \lambda_x^{'}=\lambda_x/\sqrt{2}.
\end{equation}
Implementing the expansion in Eq.\ref{eq:expansion1D} and transforming into relative ($x=x_1-x_2$) and center of mass ($X=\frac{x_1+x_2}{2}$) coordinates (similarly for y and z), Eq.\ref{eq:CoulCart} can the be described as

\begin{equation}
\begin{split}
&\sum_{n^x}^{n_1^x+n_2^x} \sum_{m^x}^{m_1^x+m_2^x} \sum_{n^y}^{n_1^y+n_2^y} \sum_{m^y}^{m_1^y+m_2^y} C_{n^x} C_{m^x} C_{n^y} C_{m^y}\\&\int dxdXdydYdzdZ\,\psi_{n^x}(x+X/2)\psi_{n^y}(y+Y/2)\Xi_{n^{z}}(z+Z/2)\frac{e^2}{4\pi\epsilon\sqrt{x^2+y^2+z^2}}\psi_{m^x}(x-X/2)\psi_{m^y}(y-Y/2)\Xi_{m^{z}}(z-Z/2),
\end{split}
\label{eq:CnSum}
\end{equation}
where $\Xi_{n^{z}}(z+Z/2) = \xi_{n_1^{z}}(z_1)\xi_{n_2^{z}}(z_1)$ and $\Xi_{m^{z}}(z-Z/2) = \xi_{m_1^{z}}(z_2)\xi_{m_2^{z}}(z_2)$.
Integrating over the center of mass coordinates and using the generating functions (Eq.\ref{eq:generator}) in the same manner described above leads us to the expression

\begin{equation}
\sum_{n^x}^{n_1^x+n_2^x} \sum_{m^x}^{m_1^x+m_2^x} \sum_{n^y}^{n_1^y+n_2^y} \sum_{m^y}^{m_1^y+m_2^y} C_{n^x} C_{m^x} C_{n^y} C_{m^y}
\sum_{l^x,p^x}^{n^x+m^x} \sum_{l^y,p^y}^{n^y+m^y}
 D_{nmlp^x} D_{nmlp^y}
 \int dxdydz\,\psi_{p^x}(x)\psi_{p^y}(y)\zeta(z)\frac{e^2}{4\pi\epsilon\sqrt{x^2+y^2+z^2}},
\label{eq:DnSum}
\end{equation}
with 
\begin{equation}
\begin{split}
 D_{nmlp} = &\frac{i^l\sqrt{2}}{\lambda}\sqrt{2^p p! \lambda \sqrt{\pi}} \frac{\sqrt{2 m! n!}}{\sqrt{2^{m+n}}} (-1)^{m+n-l} (2)^{-\frac{m+n}{2}} H_l(0)
 \sum_{q}^{n} \frac{1}{q!}(-1)^{-q}
 \frac{1}{(m+n-l+q) !}  \frac{1}{(n-q) !}  \frac{1}{(l-n+q)!} 
 \end{split},
\end{equation}
when $p=m+n-l$ and $0$ otherwise,
\begin{equation}
\zeta(z)=\int dZ\,\Xi_{n^{z}}(z+Z/2)\Xi_{m^{z}}(z-Z/2),
\end{equation}
and
\begin{equation}
\psi_p(x_i) = \frac{1}{\mathcal{N}} e^{\frac{-x_{i}^2}{2 \lambda_x^{''2}}}H_p(\frac{x_i}{\lambda_x^{''}}) ; \lambda_x^{''}=\lambda_x^{'}\sqrt{2}=\lambda_x,
\end{equation}
Eq.\ref{eq:TarnsCoul} in the main text is the same as Eq.\ref{eq:DnSum} with the difference of being represented in $r$ rather than $x$,$y$\&$z$ for the sake of a more concise description.
\newpage

%

\section{Full Configuration Interaction with larger basis set}
\label{app:MoreBasis}

As described in the main text, the exchange interaction estimates were calculated using 1 basis set element in y- and z-directions for computational speed. Table\ref{tab:MoreBasis} shows the tunnel coupling and exchange interactions calculated by diagonalizing the single and two electron Hamiltonian matrices respectively. The Hamiltonians are described using different number of basis set elements in the y-direction. As for the z-direction a separate calculation has been on the same devices with two basis set elements in that direction showing a change in exchange interaction of less than 3\% which expected given the large energy spacing ($>$15 meV) due to the strong confinement at the $\mathrm{Si/SiO_2}$ interface. The first row in Tab.\ref{tab:MoreBasis} describes the values used in the main text, which has shown convergence beyond 25 basis set elements in the x-direction. Hence, for the rows below it we use 31 basis set elements in the x-direction.

\begin{table}[h]
\caption{\label{tab:MoreBasis}%
Tunnel coupling and exchange interaction calculated using different number of basis set elements in the y-direction}
\begin{ruledtabular}
\begin{tabular}{ccccc}
\#$n^x$ &\#$n^y$&\#$n^z$&Tunnel Coupling (GHz)&Exchange (MHz)\\
\hline
51&1&1&33.92&18.28\\
31&1&1&34.18&18.46\\
31&2&1&33.50&19.11\\
31&3&1&33.88&20.24\\
31&4&1&33.93&20.56\\
31&5&1&33.93&20.69\\
\end{tabular}
\end{ruledtabular}
\end{table}

Accordingly, we confirm the validity of the results shown in the main text, as they are within a acceptable error range as compared to calculations with multiple basis set elements in all dimensions.
\newpage
\section{Noise Simulations Parameters}
\label{app:SimParam}
The simulations performed in Sec.\ref{sec:Simulations} are done under the assumption of a many electron quantum dot as per the experimental convention of charge noise characterization via single electron transistors (SETs) as charge sensors. In the many electrons regime, the behavior of SETs can be well described by the constant interaction model which involves single particle energies in addition to charging energies \cite{kouwenhoven_electron_1997}. Furthermore, in SETs with many electrons, the electron wave functions obey random matrix statistics due to any irregularities in the confining potentials in addition to electron-electron interactions and screening effects within the dot \cite{Vorojtsov_2004,beenakker_random-matrix_1997}. However, it can be safely assumed that  their wave functions would be uniformly distributed over the device area. Accordingly, for the purpose of studying the electrostatic effect of TLFs on a SET, finding the exact wave function is not necessary and a wave function that extends over the quantum dot area shall suffice. Hence, a high order eigen-function of a symmetric harmonic oscillator is used as an approximation for $\Psi_N$ (Eq.\ref{eq:Hermite}) to estimate the fluctuations described by Eq.\ref{eq_deltamu}. Given a typical 70 nm x 70 nm SET, $\Psi_N$ with $\lambda = 10$ $\mathrm{nm}$ and $nx=ny=8$ is used to represent the electron density over the dot area for the simulations presented in Sec.\ref{sec:Simulations}. 
As mentioned in the main text, the results in Ref.\cite{connors_charge-noise_2022} show a good match between measured charge noise measured using qubit spectroscopy and charge sensing which indicates that for a single electron quantum dot ($n_x=n_y=0$), on average, charge noise figures shall be in close range to the simulations presented here. We compare the distribution of the noise spectra at 1Hz for single and many electron dots as shown in Fig.\ref{fig:1NvsSET}. We observe that the mean and median values of both distributions are in close range of each other while the single electron dot shows a larger variability which further confirms the fidelity results discussed in Sec.\ref{sec:Quantum Gate Fidelities}.

\begin{figure}[h!]
\includegraphics{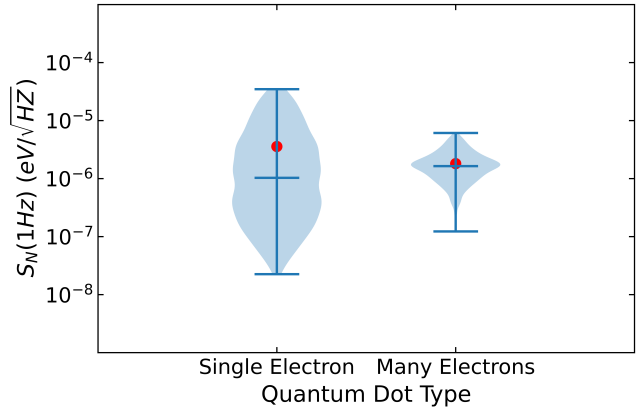}
\caption{Log-scale violin plots of the distributions of noise spectral densities at 1Hz for a single electron and many electrons quantum dots. The horizontal dashes mark the minimum, median and maximum of distributions (from bottom to top) while the red dots mark the mean values for each type of dot. \label{fig:1NvsSET} }
\end{figure}

\newpage
\section{Qubit response to whole charge fluctuations}
\label{Q_WH}
Here we study the effect of unscreened whole charge fluctuations, on single qubit frequencies and two qubit exchange interactions. The results shown below confirm the deleterious effect on qubits which was indicated in the higher noise spectral densities in Sec.\ref{sec:Simulations}.
\subsection {X-Gates}
Table.\ref{tab:EDSRWH} shows the quantum X-gate error due to coupling (transverse) noise from whole charge defects. The field due to a whole charge 50 nm away from the dot can cause a gate error of 0.26\%, while for an EDSR coupling in the y-direction (($\mathbf{E_0}=(0,E_y,0)$) a charge at the same distance (in the y-direction) would induce errors less than $10^{-5} \%$. Such results demonstrate the confinement effect on the sensitivity of quantum gate performance to charge noise.

\begin{table}[h]
\caption{\label{tab:EDSRWH}%
Coupling Noise induced X-Gate errors is the presence of one whole charge TLF}
\begin{ruledtabular}
\begin{tabular}{ccddd}
TLF&(x,y,z)\footnote{The single quantum dot is centered at (0,0,0) and is 2 nm below the $\mathrm{Si/SiO_2}$ interface} nm&
\multicolumn{1}{c}{\textrm{Gate Error \%}}\\
\hline
WH&(50,0,2)&0.26\\
WH&(0,50,2)&<$$10^{-5}$$\\
WH&(100,0,2)&0.001\\
\end{tabular}
\end{ruledtabular}
\end{table}

We further study the effect of unscreened whole charges as a source of dephasing noise. Table.\ref{tab:EDSR_dephWH} shows the frequency shift and quantum gate error due to a single whole charge defect. A trap located 50 nm away to side of the qubit (at the edge of the 2DEG) can shift the qubit frequency by 5.78 MHZ, which attenuates and shifts the Rabi oscillations to a regime that cannot be effectively compared to the ideal operation and is a further indication for the dismissal of un-screened whole charge fluctuations as a source of noise in this model.

\begin{table}[h!]
\caption{\label{tab:EDSR_dephWH}%
Dephasing noise induced qubit Frequency shifts and X-Gate errors is the presence of one TLF}
\begin{ruledtabular}
\begin{tabular}{cccc}
TLF&(x,y,z)\footnote{The single quantum dot is centered at (0,0,0) and is 2 nm below the $\mathrm{Si/SiO_2}$ interface} nm& $\Delta f (Hz)$&
\multicolumn{1}{c}{\textrm{Gate Error \%}}\\
\hline
WH&(0,50,2)&2.66*$10^5$&0.48 \\
WH&(50,0,2)&5.78*$10^6$& -- \\
\end{tabular}
\end{ruledtabular}
\end{table}

\subsection{SWAP Gates}

Under the assumption of a depleted DQD region, an unscreened whole charge is positioned 100 nm away from center of the DQD, where the 2DEG is located shows a 25\% shift in exchange interaction.

\begin{table}[h]
\caption{\label{tab:ExchangeWH}%
Shift in the exchange interaction calculated using a full configuration interaction model in the presence of TLFs in close vicinity of the DQD system}
\begin{ruledtabular}
\begin{tabular}{ccddd}
TLF&(x,y,z) nm\footnote{The DQD is centered at (0,0,0) and is 2 nm below the $\mathrm{Si/SiO_2}$ interface}&
\multicolumn{1}{c}{\textrm{$\Delta J/J$}}\\
\hline
WH&(100,0,2)&0.25\\
\end{tabular}
\end{ruledtabular}
\end{table}

Furthermore the SWAP gate fidelity at a trap density of ($\mathrm{10^{11}cm^{-2}}$) with uniform spatial and frequency distribution is calculated as in Sec.\ref{sec:Quantum Gate Fidelities} showing a fidelity of $84.9\%$ when compare to the ideal operation.

\begin{table}[h]
\caption{\label{tab:ExchangeFWH}%
SWAP Gate fidelity in the presence of TLFs distrubutions}
\begin{ruledtabular}
\begin{tabular}{ccddd}
TLF&Density ($\mathrm{cm^{-2}}$)&
\multicolumn{1}{c}{\textrm{Fidelity \%}}\\
\hline
WH&$10^{11}$&84.9\\
\end{tabular}
\end{ruledtabular}
\end{table}

\clearpage

\section{Quantum Gate fidelities vs TLF density}
\label{sec:FidelityVsDensity}

\begin{table*}[h]
\caption{\label{tab:XGatevsDensity}
One-qubit X-Gate fidelity at different DWP densities.}
\begin{ruledtabular}
\begin{tabular}{cccccc}
Density($\mathrm{cm^{-2}}$)&Avg. No.of DWPs&Simulated Area&Spatial Dist.&Frequency Dist.&Average Fidelity (\%)
\\ \hline
 \multirow{4}{*}{$10^{9}$}&\multirow{4}{*}{$25$}&\multirow{4}{*}{$1600$ $  \mathrm{nm}*1600$ $\mathrm{ nm}$}&\multirow{2}{*}{1}&A&99.99997\\
 &&&&B&99.9999(7)\\
 &&&\multirow{2}{*}{2}&A&100\\
 &&&&B&99.999986\\
 \hline
  \multirow{4}{*}{$5*10^{9}$}&\multirow{4}{*}{$32$}&\multirow{4}{*}{$800$ $\mathrm{nm}*800$ $\mathrm{nm}$}&\multirow{2}{*}{1}&A&99.9995\\
 &&&&B&94.7762\\
 &&&\multirow{2}{*}{2}&A&99.99999\\
 &&&&B&99.99999\\
 \hline
  \multirow{4}{*}{$10^{10}$}&\multirow{4}{*}{$16$}&\multirow{4}{*}{$400$ $\mathrm{nm}*400$  $\mathrm{nm}$}&\multirow{2}{*}{1}&A&99.9961\\
 &&&&B&99.9972\\
 &&&\multirow{2}{*}{2}&A&94.0606\\
 &&&&B&97.5359\\
 \hline
  \multirow{4}{*}{$5*10^{10}$}&\multirow{4}{*}{$20$}&\multirow{4}{*}{$200$ $\mathrm{nm}*200$ $\mathrm{nm}$}&\multirow{2}{*}{1}&A&99.5500\\
 &&&&B&99.3100\\
 &&&\multirow{2}{*}{2}&A&99.1000\\
 &&&&B&99.0400\\
 \hline
  \multirow{4}{*}{$10^{11}$}&\multirow{4}{*}{$40$}&\multirow{4}{*}{$200$ $\mathrm{nm}*200$ $\mathrm{nm}$}&\multirow{2}{*}{1}&A&99.9990\\
 &&&&B&99.5100\\
 &&&\multirow{2}{*}{2}&A&97.5500\\
 &&&&B&99.1000\\
 \hline
  \multirow{4}{*}{$5*10^{11}$}&\multirow{4}{*}{$50$}&\multirow{4}{*}{$100$ $\mathrm{nm}*100$ $\mathrm{nm}$}&\multirow{2}{*}{1}&A&98.1700\\
 &&&&B&80.9300\\
 &&&\multirow{2}{*}{2}&A&98.4900\\
 &&&&B&94.5400\\
 \hline
  \multirow{4}{*}{$10^{12}$}&\multirow{4}{*}{$100$}&\multirow{4}{*}{$100$ $\mathrm{nm}*100$ $\mathrm{nm}$}&\multirow{2}{*}{1}&A&96.8900\\
 &&&&B&93.3000\\
 &&&\multirow{2}{*}{2}&A&95.0800\\
 &&&&B&93.9000\\
\end{tabular}
\end{ruledtabular}
\end{table*}
\newpage
\begin{table*}
\caption{\label{tab:SWAPGatevsDensity}
Two-qubit SWAP-Gate fidelity at different DWP densities.}
\begin{ruledtabular}
\begin{tabular}{cccccc}
Density($\mathrm{\mathrm{cm^{-2}}}$)&Avg. No.of DWPs&Simulated Area&Spatial Dist.&Frequency Dist.&Average Fidelity (\%)
\\ \hline
 \multirow{4}{*}{$10^{9}$}&\multirow{4}{*}{$25$}&\multirow{4}{*}{$1600$ $\mathrm{nm}*1600$ $\mathrm{nm}$}&\multirow{2}{*}{1}&A&99.99843\\
 &&&&B&99.99843\\
 &&&\multirow{2}{*}{2}&A&99.99995\\
 &&&&B&100\\
 \hline
  \multirow{4}{*}{$5*10^{9}$}&\multirow{4}{*}{$32$}&\multirow{4}{*}{$800$ $\mathrm{nm}*800$ $\mathrm{nm}$}&\multirow{2}{*}{1}&A&99.99999\\
 &&&&B&99.99999\\
 &&&\multirow{2}{*}{2}&A&88.81690\\
 &&&&B&99.99997\\
 \hline
  \multirow{4}{*}{$10^{10}$}&\multirow{4}{*}{$16$}&\multirow{4}{*}{$400$ $\mathrm{nm}*400$ $\mathrm{nm}$}&\multirow{2}{*}{1}&A&99.9999\\
 &&&&B&99.99935\\
 &&&\multirow{2}{*}{2}&A&99.99758\\
 &&&&B&91.58767\\
 \hline
  \multirow{4}{*}{$5*10^{10}$}&\multirow{4}{*}{$20$}&\multirow{4}{*}{$200$ $\mathrm{nm}*200$ $\mathrm{nm}$}&\multirow{2}{*}{1}&A&97.52754\\
 &&&&B&99.91836\\
 &&&\multirow{2}{*}{2}&A&98.10858\\
 &&&&B&97.39445\\
 \hline
  \multirow{4}{*}{$10^{11}$}&\multirow{4}{*}{$40$}&\multirow{4}{*}{$200$ $\mathrm{nm}*200$ $\mathrm{nm}$}&\multirow{2}{*}{1}&A&99.58308\\
 &&&&B&96.35075\\
 &&&\multirow{2}{*}{2}&A&91.74441\\
 &&&&B&98.02771\\
 \hline
  \multirow{4}{*}{$5*10^{11}$}&\multirow{4}{*}{$50$}&\multirow{4}{*}{$100$ $\mathrm{nm}*100$ $\mathrm{nm}$}&\multirow{2}{*}{1}&A&84.05879\\
 &&&&B&97.05879\\
 &&&\multirow{2}{*}{2}&A&90.43929\\
 &&&&B&96.37993\\
 \hline
  \multirow{4}{*}{$10^{12}$}&\multirow{4}{*}{$100$}&\multirow{4}{*}{$100$ $\mathrm{nm}*100$ $\mathrm{nm}$}&\multirow{2}{*}{1}&A&69.34702\\
 &&&&B&77.96889\\
 &&&\multirow{2}{*}{2}&A&96.65003\\
 &&&&B&53.95905\\
\end{tabular}
\end{ruledtabular}
\end{table*}
Note that the simulated area has been adapted to TLFs density to make sure the number of TLFs present is enough to allow for a reasonable uniform spatial distribution.

\section{Quantum Gate fidelities vs Noise Spectral Densities}
\label{app:FidelityVsSpectra}

As shown in Fig.\ref{fig:X-GateF_Den} and Fig.\ref{fig:SWAP-GateF_Den} of the main text, there is a large variability of quantum gate fidelities over a single TLF density. Such a variability can explained by two factors. First, as discussed in the main text, the qubits are most sensitive to fluctuations induced by the TLFs in their close vicinity. Hence, the qubit fidelity is subject to the state and activity of such TLFs during the quantum gate operations in spite of their large contribution to the noise spectral density. Second the distribution of noise spectra at certain TLF density as shown in Fig.\ref{fig:PSD_all} and Fig.\ref{fig:DensityVsNoise} can also contribute to the variability of quantum gate fidelities. Nonetheless, one can still expect a statistical correlation between noise figures and quantum gate fidelities, which can confirm the relevance of such noise figures as metric for device quality. Given the time cost of realistic time-dependent simulations of quantum gates in the presence of noise, we are not able to detect such correlations with the simulations presented in Sec.\ref{sec:Quantum Gate Fidelities}. However, in this appendix we study the correlation between noise spectral densities and quantum gate performance by adopting a simpler approach for calculating quantum gate fidelity. 100 different random distributions (spatially and temporally) of DWPs are created at a density of $\mathrm{10^{11} cm^{-2}}$. Fig.\ref{fig:PSDs_100} shows the mean, minimum and maximum of the calculated spectral densities with an inset showing the distribution of the spectral densities at 1Hz.  Fig.\ref{fig:PSDs_100}(a) describes the electrostatic potential noise (N) as described in Sec.\ref{sec:Simulations} while  Fig.\ref{fig:PSDs_100}(b) describes the noise on the qubit frequency (NQ) calculated using the electric field due to the TLFs as described in Sec.\ref{sub_OneQubit}. We calculate average X-gate fidelities over 1000 gate operations per TLF distribution. However, we assume that switching events of TLFs occur only between (not during) gate operations which allows for an analytic calculation of the fidelity. In other words, we assume a constant frequency  shift throughout each X- gate operation which can change from operation to another according to the TLFs switching frequencies.

\begin{figure}[h!]
\includegraphics{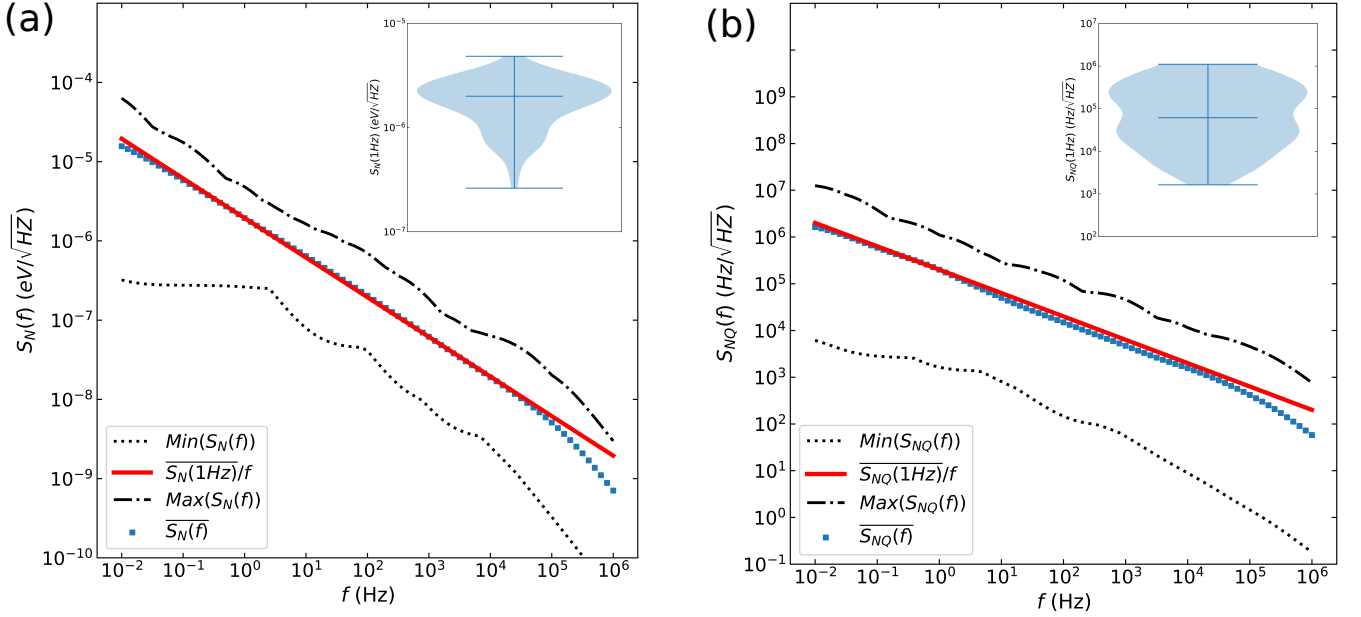}
\caption{Log-scale plot of the (a) electrostatic potential and (b) qubit frequency noise spectral densities calculated for 100 random distributions of DWPs. The regions between the dotted and dashed lines span the range from the minimum and maximum calculated values at each frequency, the blue squares mark the mean spectral density at every frequency and the red lines represent $A/\sqrt{f}$ where A is the mean spectral density at 1Hz. The inset to the figure shows a violin plot of the distribution of the spectral densities at 1Hz.} 
\label{fig:PSDs_100}
\end{figure}

Single qubits gates are sensitive to the qubit frequency shifts which in turn are dependent on the electric fields from TLFs. Such electric field dependence introduces a third variability factor when studying correlations between electrostatic noise spectral densities, which quantifies potential noise, and one-qubit gate fidelities. While strong electrostatic potentials naturally imply a strong electric field, this is not necessarily true when studying spin in quantum dots. This can observed by comparing the distributions in Fig.\ref{fig:PSDs_100}(a) and Fig.\ref{fig:PSDs_100}(b) which are calculated for the same TLF configurations. To elaborate, we can compare the effect of two TLFs , one located directly above the center of a quantum dot (TLF1) and another that is located to its left (TLF2) (see first and third rows of Table.\ref{tab:EDSR_deph}). TLF1 would have a stronger effect on electrostatic potential of the dot, however, a weaker effect on the qubit frequency due to the confinement effects. Furthermore, two TLFs located on opposite sides of the quantum dot would have zero electric field in the direction of their separation, while electrostatic potential would be the sum of both.  However, as shown in Fig.\ref{fig:PSDs_100}(b) and Ref.\cite{yoneda_quantum-dot_2018}, noise spectral densities extracted from qubit dynamics are described as noise on the qubit frequencies (NQ), which eliminates the third variability factor discussed above. Furthermore, we extract a monotonic relationship between the mean electrostatic noise  spectral density ($\overline{S_N(1Hz)}/f$) and mean qubit frequency noise spectral density  ($\overline{S_{NQ}(1Hz)}/f$) as shown in Fig.\ref{fig:PSDsVSPSDsF}.

\begin{figure}[h!]
\includegraphics{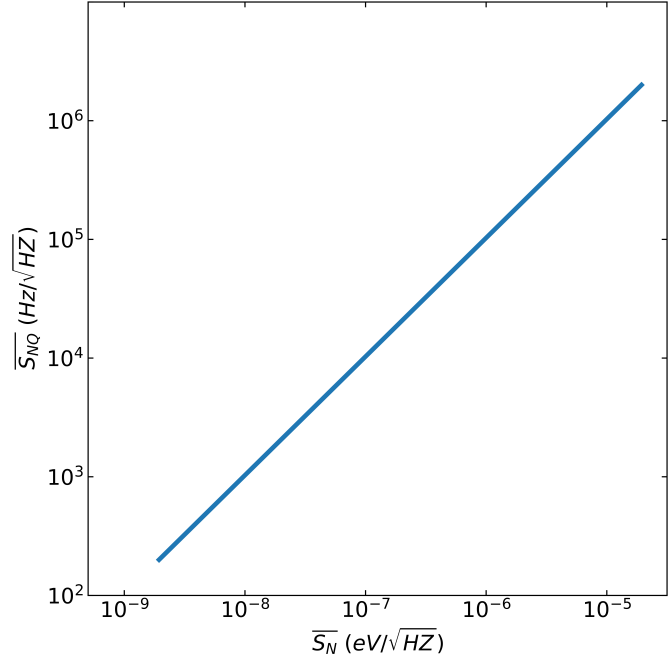}
\caption{A Log-scale plot of the mean spectral density of electrostatic noise ($S_N$) vs spectral density on the qubit frequency ($S_{NQ}$) showing a linear monotonic relationship} 
\label{fig:PSDsVSPSDsF}
\end{figure}

Under the assumption that qubit frequency shifts occur only between different gate operations, the average X-gate fidelity can be described by 

\begin{equation}
F=\frac{1}{N_X} \sum_i^{N_X} \frac{\omega_d^2}{\omega_d^2+\Delta\omega_i^2} sin^2(\frac{t_X\sqrt{\omega_d^2+\Delta\omega_i^2}}{2}),
\end{equation}
with $\omega_d = 0.5 g \mu_b B_c(X(t))$ , $\Delta\omega_i$ is the total qubit frequency shift due to TLFs during the $i^{th}$ gate operation, $t_X$ is the X-gate time defined in Sec.\ref{sub_OneQubit} and $N_X$ is the total number of gate operations which is 1000 in this study.

Fig.\ref{fig:PSDsF_100vsFidelity} shows the X-gate infidelity with respect to the spectral density at 1Hz of the noise on the qubit frequency.  Due to the aforementioned variability, a distribution of infidelity is present for any given range of qubit frequency noise. The region between the dotted and dashed line in Fig.\ref{fig:PSDsF_100vsFidelity} span the range from the minimum and maximum infidelities and the blue line is a linear fit obtained with a linear correlation coefficient (coefficient of determination) of 0.433 indicating a power-law relationship.

\begin{figure}[h!]
\includegraphics{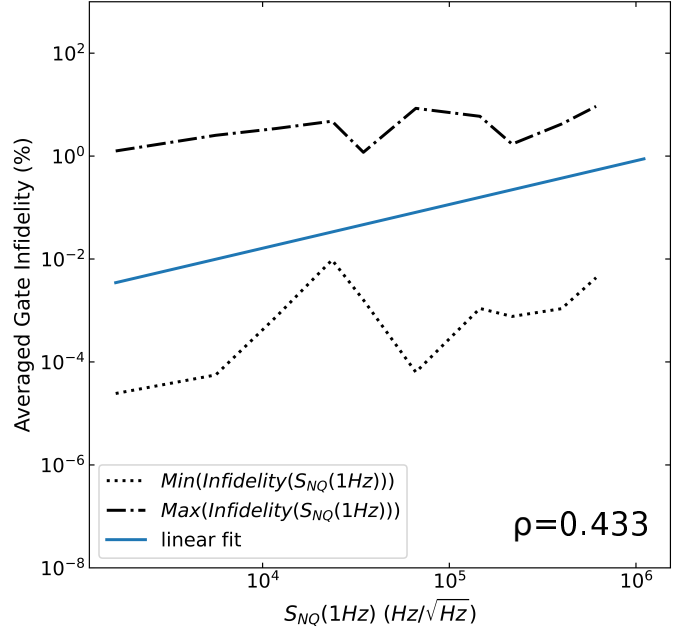}
\caption{Log-scale plot of the noise spectral density on the qubit frequency vs the averaged X-gate infidelity for 100 different random distributions of DWPs. The correlation coefficient ($\rho$) is found to be 0.433} 
\label{fig:PSDsF_100vsFidelity}
\end{figure}

\newpage
\end{widetext}

\newpage
\bibliography{References}

\end{document}